\documentclass[10pt,journal,compsoc]{IEEEtran}
\usepackage{array}
\usepackage{amsmath}
\usepackage{amssymb}
\usepackage{graphicx}
\usepackage{multirow}
\usepackage{subfigure}
\usepackage{epstopdf}
\usepackage{eqparbox}
\usepackage{cite}
\usepackage{url}
\begin{document}
\title{Personalized Video Recommendation Using Rich Contents from Videos}
\author{
  Xingzhong Du, Hongzhi Yin*, Ling Chen, Yang Wang, Yi Yang, Xiaofang Zhou
  \thanks{Xingzhong Du, Hongzhi Yin, and Xiaofang Zhou are with School of Information Technology \& Electric Engineering, The University of Queensland, Australia.(e-mail: x.du@uq.edu.au; h.yin1@uq.edu.au; zxf@itee.uq.edu.au). Hongzhi Yin is the corresponding author (denoted by *).}
  \thanks{Ling Chen and Yi Yang are with the Centre for Artificial Intelligence, University of Technology Sydney, Australia.(e-mail:ling.chen@uts.edu.au; yi.yang@uts.edu.au)}
  \thanks{Yang Wang is with School of Computer Science \& Engineering, The University of New South Wales, Australia.(e-mail: wangy@cse.unsw.edu.au).}
}
\IEEEtitleabstractindextext{%
  \begin{abstract}
  Video recommendation has become an essential way of helping people explore the massive videos and discover the ones that may be of interest to them. In the existing video recommender systems, the models make the recommendations based on the user-video interactions and single specific content features. When the specific content features are unavailable, the performance of the existing models will seriously deteriorate. Inspired by the fact that rich contents (e.g., text, audio, motion, and so on) exist in videos, in this paper, we explore how to use these rich contents to overcome the limitations caused by the unavailability of the specific ones. Specifically, we propose a novel general framework that incorporates arbitrary single content feature with user-video interactions, named as collaborative embedding regression (CER) model, to make effective video recommendation in both in-matrix and out-of-matrix scenarios. Our extensive experiments on two real-world large-scale datasets show that CER beats the existing recommender models with any single content feature and is more time efficient. In addition, we propose a priority-based late fusion (PRI) method to gain the benefit brought by the integrating the multiple content features. The corresponding experiment shows that PRI brings real performance improvement to the baseline and outperforms the existing fusion methods.
  \end{abstract}
  
  % Note that keywords are not normally used for peerreview papers.
  \begin{IEEEkeywords}
    Recommender model, personalization, video content analysis, rich content features, late fusion.
  \end{IEEEkeywords}
}

\maketitle

\IEEEdisplaynontitleabstractindextext

\section{Introduction}
Watching online videos has become one of the indispensable entertainment activities in daily life. Many famous websites, such as YouTube, Netflix and Hulu, host a tremendous number of videos to meet such demand. The massive video repositories have placed an enormous burden on users when trying to find videos of interest\cite{DBLP:journals/tkde/AdomaviciusT05}\cite{DBLP:reference/sp/2015rsh}. To address this problem, most video websites have adopted recommender systems as a promising way to help users explore the world of videos\cite{DBLP:conf/recsys/DavidsonLLNVGGHLLS10}\cite{gomez2015netflix}. Existing recommender methods can be categorized into three classes\cite{DBLP:journals/tkde/AdomaviciusT05}: content-based, collaborative filtering (CF)-based, and hybrid. Content-based methods \cite{DBLP:journals/tkde/QianFZM14} recommend items to users based on the content similarities between the user profile and item contents. CF-based methods \cite{DBLP:journals/tkde/BuSXCHC16,DBLP:journals/tist/WangYCSSZ17} accomplish the same task by the behavior similarities between the users or items. Hybrid methods\cite{Agarwal:2009:RLF:1557019.1557029}\cite{Hu:2013:PRV:2488388.2488441} seek the best of both worlds by combining both content and CF-based methods, and have gained increasing popularity in recent years \cite{DBLP:conf/kdd/WangB11}\cite{DBLP:conf/kdd/WangWY15}\cite{DBLP:conf/kdd/ZhangYLXM16}. 

Most existing hybrid recommender models \cite{DBLP:conf/kdd/WangB11}\cite{DBLP:conf/kdd/WangWY15}\cite{DBLP:conf/kdd/ZhangYLXM16} use only textual contents to facilitate recommendations. When these models are used for video recommendation, they are fragile because the textual contents are often missing, scarce, or poor-quality for user-generated videos. For example, plenty of videos on Youtube only have titles, and many of them 
are not suitable for the content-based recommendation task as they are deliberately titled to attract users. If the hybrid recommender models are forced to generate recommendations with these texts, their performance will be much worse than the expectation. The limitation caused by the unavailability of the textual contents has been noticed in a few works \cite{DBLP:conf/civr/YangMHYYL07}\cite{DBLP:conf/nips/OordDS13}\cite{DBLP:journals/jodsn/DeldjooECGPQ16} for video, music or product recommendation. They propose to exploit non-textual content features (e.g., color histograms) to overcome the limitation. However, similar to models using textual contents, all these models  \cite{DBLP:conf/civr/YangMHYYL07}\cite{DBLP:conf/nips/OordDS13}\cite{DBLP:journals/jodsn/DeldjooECGPQ16} only explore single specific non-textual content features. Thus, these models still face the performance drop when their dependent content features are unavailable.

The recent study in \cite{DBLP:journals/tkde/HoilesAK17} shows that multiple types of content features can influence users' choices on videos. Enlightened by this new finding, we propose to use rich contents from videos to enhance the recommendation task. Our rich content features consist of both textual content features and non-textual content features. The textual content features are made up of video descriptions and video meta information such as actors and directors, and the non-textual content features are made up of audios \cite{aly2013axes}, scenes \cite{DBLP:journals/corr/SimonyanZ14a} and motions \cite{DBLP:conf/iccv/WangS13a} from videos themselves. With rich content features, our goals are two folds: (1) we want to discover a general and effective hybrid model, which is able to integrate any single content feature into collaborative filtering, to marginalize the performance drop caused by the unavailability of one specific content; (2) we want to explore a multiple feature fusion method, which is inspired by the success of the multiple feature fusion in other relevant areas \cite{aly2013axes}\cite{DBLP:conf/cvpr/XuYH15}, to further improve the recommendation accuracy .

With the two-fold goals, we start our study by analyzing the performance of existing hybrid models \cite{DBLP:conf/kdd/WangB11}\cite{DBLP:conf/nips/OordDS13}\cite{DBLP:conf/kdd/WangWY15}\cite{DBLP:conf/aaai/HeM16} with single content features, because each of them is tightly coupled with one specific content feature. The performance of these existing models are tested in two different but important scenarios \cite{DBLP:conf/kdd/WangB11}: in-matrix recommendation and out-of-matrix recommendation. These two scenarios are also known as warm-start and cold-start recommendations. Specifically, for a test case, if the video in it has appeared in training data (i.e., the user-video interaction matrix), then it is a in-matrix recommendation; otherwise, it is a out-of-matrix recommendation. We find that none of these existing models could achieve very sound performance in both in-matrix and out-matrix scenarios even with their original content features. In particular, we observe that weighted matrix factorization (WMF)-based models \cite{DBLP:conf/icdm/HuKV08}\cite{DBLP:conf/kdd/WangB11}\cite{DBLP:conf/nips/OordDS13}\cite{DBLP:conf/kdd/WangWY15} achieve higher accuracy than Bayesian personalized ranking (BPR)-based models \cite{DBLP:conf/uai/RendleFGS09} \cite{DBLP:conf/aaai/HeM16} in the in-matrix scenario, but lower accuracy in the out-of-matrix scenario. WMF-based models' poorer performance in out-of-matrix scenario is attributed to that their model are designed for in-matrix recommendation especially. For example, collaborative deep learning (CDL) model is a WMF-based model proposed in \cite{DBLP:conf/kdd/WangWY15}. CDL uses stack denoising auto-encoder (SDAE) to make its content side serve in-matrix recommendation when ratings are sparse. Accordingly, CDL neglects its content side is also important for out-of-matrix recommendation. To address that, we propose a collaborative embedding regression (CER) model to effectively incorporate collaborative filtering (CF) with single content features. Our extensive evaluations show that, CER significantly outperforms both WMF and BPR-based models in the out-of-matrix scenario with any content feature, while keeps the excellent performance of WMF-based models in the in-matrix scenario. Moreover, CER's model training is more efficient on large datasets than the other models'.

Most existing hybrid recommender models including our proposed CER model are designed to work with single content features, which neglects the performance improvement from the multiple feature fusion. Thus, how to design an effective fusion method to further improve the recommendation accuracy is another challenge in our study. In other areas, there are two widely used yet independent strategies to fuse multiple content features\cite{Cui:2010:MFF:1807167.1807216}: early fusion and late fusion. The early fusion combines the multiple content features before model training \cite{aly2013axes}. Even though sometimes early fusion methods obtain better performance than the individual content features, they have a number of limitations. First, the input content feature space is fixed , so early fusion models need to retrain from a scratch when different content features come. This property makes early fusion unadaptive to the streaming data from the real-world recommender systems. Second, 
Most works on early fusion concatenate multiple content features into single ones as the input for model training \cite{srivastava2012learning}\cite{DBLP:conf/kdd/ZhangYLXM16}\cite{Wang:2014:EMR:2732296.2732301}. The concatenation results into very high dimension that leads to extremely unprecedentedly computational costs in training. Third, another line of the early fusion methods make all the content features additive in a homogeneous latent space and use the sum of them as the input for training \cite{DBLP:conf/kdd/ZhangYLXM16}. However, the textual, audio, visual and motion information contained in videos are quite diverse and heterogeneous. It is unreasonable to add them in a homogeneous latent space. The other line of the fusion research focuses on the late fusion of multiple features. A typical late fusion method obtains the fused scores by the weighted sums of the scores from different content features \cite{DBLP:journals/ftir/Liu09}. Learning-to-rank techniques (e.g., ranking SVM) \cite{DBLP:journals/ftir/Liu09}\cite{aly2013axes} are state-of-the-art late fusion methods. Compared to early fusion, late fusion treats the content features in heterogeneous space and is more flexible to adapt to the streaming data and more efficient in training \cite{DBLP:conf/icml/CaoQLTL07}. A number of recent successful multimedia event detection systems \cite{aly2013axes}\cite{DBLP:conf/cvpr/XuYH15} have adopted late fusion in combining multiple features. 

In this paper, we explore the late fusion strategy to further improve CER using rich content features. Specifically, we propose a priority-based late fusion method (PRI) whose innovation is using exponential weights to model the overwhelming influences from the stronger content features. Given rich content features, PRI firstly prioritizes the content features by evaluating their recommendation accuracy in validation. After that, PRI applies grid search to obtain the optimal base for the exponential weights. Finally, PRI calculates each content feature's weight using the optimal base and the exponents derived from the priority positions. The effectiveness of PRI has been evaluated by comparing it with five fusion methods including both early and late fusion on two real-world large-scale datasets. 

To summarize, the contributions of this paper include:
\begin{itemize}
	\item We have proposed a novel and versatile framework for effective personalised video recommendation in both out-of-matrix and in-matrix scenarios by exploiting rich content features from videos.
	\item We have deeply analyzed and studied the performance of existing hybrid models in both the in-matrix and out-of-matrix scenarios. Based on the sound study, we propose a collaborative embedding regression (CER) model, which effectively combines collaborative filtering (CF) with arbitrary single textual or non-textual content feature, to generate more accurate video recommendation in both in-matrix and out-of-matrix scenarios.
	\item We further study how to generate more accurate recommendations for new videos by fusing rich content features. We find existing fusion methods, either early or late, cannot consistently make performance improvement due to large accuracy divergences between different content features. We therefore propose a priority-based late fusion (PRI) method that prioritizes the content features and assigns the exponential weights according to the priorities. 
\end{itemize}

\section{Background and Related Work}
\begin{figure}[!htb]
	\centering
	\includegraphics[width=7cm]{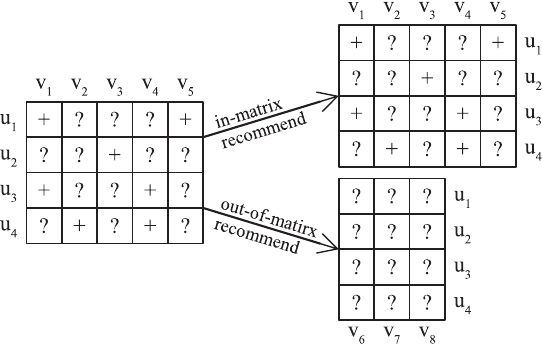}
  \vspace{-5pt}
	\caption{Implicit rating matrix, in-matrix recommendation, and out-of-matrix recommendation}
  \label{fig:iomatrix}
  \vspace{-20pt}
\end{figure}

Given $m$ users and $n$ items, $r_{ij}\in{}\{?, +\}$ denotes the $i^{th}$ user's implicit feedback on the $j^{th}$ item, where `$+$' means $i^{th}$ user likes $j^{th}$ item; `$?$' means $i^{th}$ user dislikes or is not aware of $j^{th}$ item. As a convention\cite{DBLP:conf/uai/RendleFGS09}, we transform $r_{ij}$ to $\{0, 1\}$ as a implicit rating. Putting all the ratings together forms the implicit rating matrix denoted as $R=\{0,1\}^{m\times{}n}$ for recommender models, as shown in Fig.~\ref{fig:iomatrix}.  

Given a target user, a recommender system is required to find personalized top-$k$ items that the user is potentially interested in.
The task can be further divided into two scenarios, namely in-matrix and out-of-matrix. In the in-matrix scenario, systems recommend top-$k$ items which have not been rated by the target user but have been rated by other users \cite{DBLP:conf/recsys/CremonesiKT10} \cite{DBLP:conf/recsys/DavidsonLLNVGGHLLS10}. Based on the collaborations between similar users or items, state-of-the-art models \cite{DBLP:conf/icdm/HuKV08}\cite{DBLP:conf/kdd/WangB11}\cite{DBLP:conf/nips/OordDS13}\cite{DBLP:conf/kdd/WangWY15} apply collaborative filtering (CF) to generate recommendations. In the out-of-matrix scenario, systems recommend top-$k$ new items that have not been rated by any user \cite{DBLP:conf/kdd/WangB11} (i.e., cold-start recommendation \cite{DBLP:conf/kdd/WangB11}\cite{DBLP:journals/tkde/ZhaoLHCWL16}). Since the collaborations between users or items are not existing, CF-based models become ineffective, whereas content-based models perform better.

\begin{figure*}[!hbt]
  \centering
  \includegraphics[width=16cm]{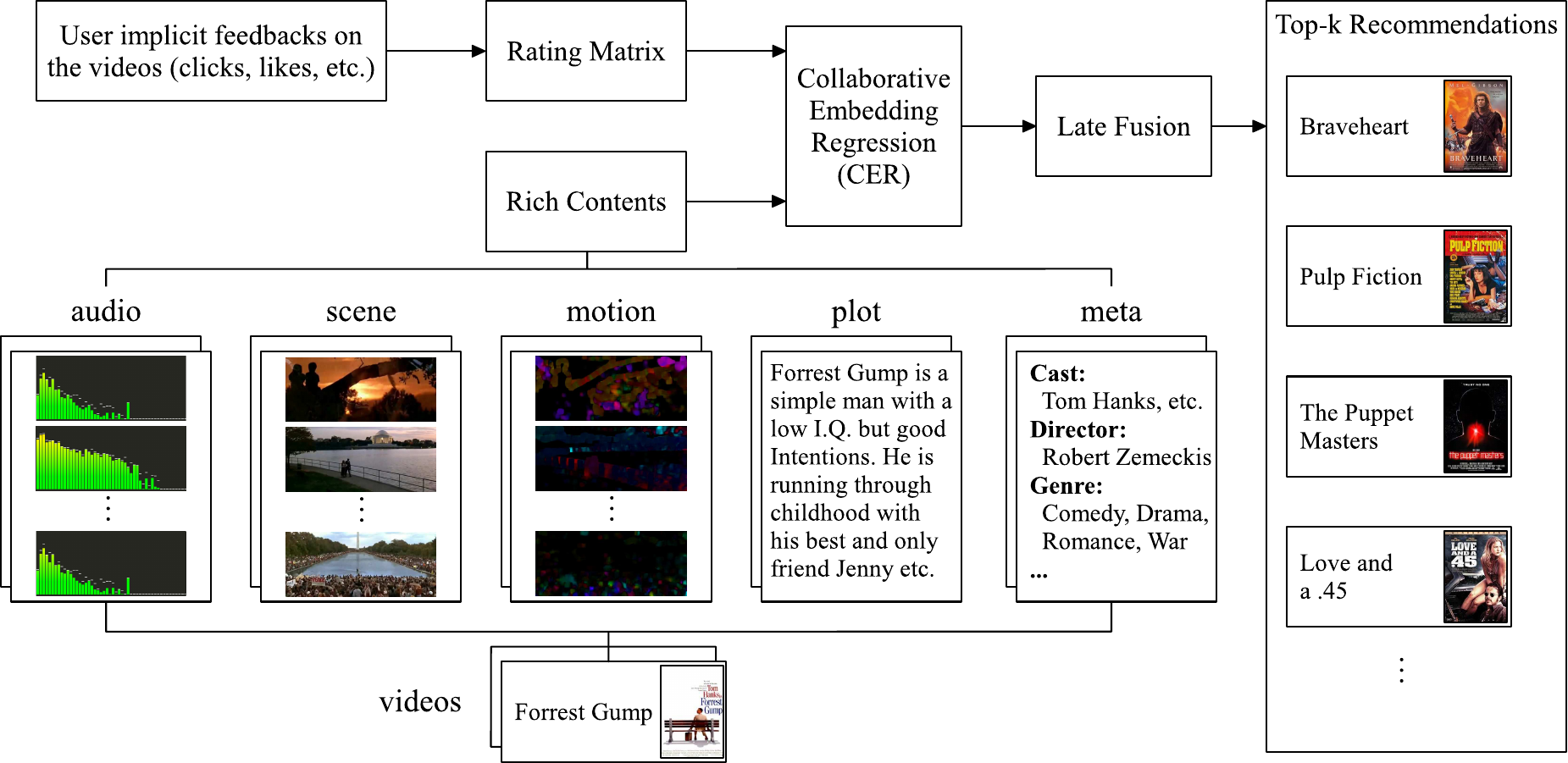}
  \vspace{-5pt}
  \caption{A running example to illustrate how the rich content features are extracted from the videos and used with the user implicit ratings to generate the personalized video recommendation.}
  \label{fig:flowchart}
  \vspace{-20pt}
\end{figure*}

Weighted matrix factorization (WMF)\cite{DBLP:journals/computer/KorenBV09} and Bayesian personalized ranking (BPR)\cite{DBLP:conf/uai/RendleFGS09} represent the state-of-the-art recommender models in in-matrix scenario.
Both of them construct objective functions based on matrix factorization and learn collaborations between users or items during the training. After training, a latent vector is learned for each user or item. Users' ratings on items are then predicted by the inner products between latent vectors. Finally, items with the highest predicted ratings will be selected as the top-$k$ recommendation for users.
The major difference between WMF and BPR is the learning objective. In particular, WMF \cite{DBLP:journals/computer/KorenBV09} aims at minimizing rating prediction errors, while BPR \cite{DBLP:conf/uai/RendleFGS09} aims at preserving pair-wise personalized orders.

Recently, both WMF and BPR were extended to incorporate content features, so their variants can be applied to generate recommendations in both in-matrix and out-of-matrix scenarios.
The representative WMF-based models include collaborative topic regression (CTR)\cite{DBLP:conf/kdd/WangB11}, deep content-based music recommender model (DPM)\cite{DBLP:conf/nips/OordDS13} and collaborative deep learning (CDL)\cite{DBLP:conf/kdd/WangWY15}. CTR and CDL are specific to the textual features of items, while DPM is specific to audio features of items. The representative BPR-based models are visual Bayesian personalized ranking (VBPR)\cite{DBLP:conf/aaai/HeM16} and Visual-CLiMF\cite{DBLP:conf/recsys/RoyG16}. VBPR and Visual-CLiMF are specific to visual features from pretrained convolutional neural networks (CNN). Visual-CLiMF enhances VBPR by learning the approximate reciprocal rank instead of pair-wise rank in the optimization. Different from above models, VideoTopic proposed in \cite{DBLP:conf/ism/ZhuSW13} use textual and visual features simultaneously to perform out-of-matrix video recommendations, which is actually a kind of early fusion. Because poor visual feature and few user collaborations are used in model training, VideoTopic's accuracy is much lower than those achieved by CF-based models \cite{DBLP:conf/kdd/WangB11} \cite{DBLP:conf/kdd/WangWY15} \cite{DBLP:conf/aaai/HeM16}.

\section{Video Content Features}
The whole work flow in our study that covers from rich content feature extraction to personalized video recommendation generation is depicted in Fig. \ref{fig:flowchart}. In this section, we firstly describe how the rich content features, including both textual and non-textual, are extracted for video recommendation. These features are crucial for the training of the hybrid models, and they directly decide the out-of-matrix recommendation performance.

\subsection{Textual Content Features}

Many existing hybrid video recommender systems\cite{DBLP:conf/recsys/DavidsonLLNVGGHLLS10}\cite{DBLP:conf/kdd/WangWY15}\cite{gomez2015netflix} make aware of the video contents by texts, which include but not limit to titles, descriptions, reviews as well as meta information for the videos. Based on these texts, two kinds of textual content features are often extracted: word vectors and meta vectors. In our work, we construct one word vector and one meta vector for each video. Word vectors are extracted from the titles, descriptions and reviews. The process concatenates texts into single documents, removes stop words, and collects tokens by stemming \cite{DBLP:conf/kdd/WangWY15}. After that, a number of top tokens of the highest TF-IDF values are selected to form the vocabulary. With the vocabulary, a word vector is created for each video with token frequencies. 
Meta vectors are extracted from official information about videos such as producers, countries, languages, release dates, actors, genres and so on. The process selects a number of top meta tokens according to the highest global frequencies to form the vocabulary. After that, a binary meta vector is created for each video according to token existences because a meta token for a video appears at most once. 

\subsection{Non-Textual Content Features}
In addition to textual contents, videos themselves also contain non-textual contents. Yang et. al\cite{DBLP:conf/civr/YangMHYYL07} extract the normalized color histogram and aural tempos as non-textual content features for videos. However, the experimental results reported in\cite{DBLP:conf/civr/YangMHYYL07} show that these features are not effective for video recommendation. One possible reason is that these features fail to distinguish between videos that share similar colors but have irrelevant contents. For example, given a video about the sky and another video about the sea, the normalized color histogram will result in a high similarity between the two videos due to the common color blue. In this case, hybrid models would recommend sky-related videos to users who like seas in a high probability.

The limitations of the non-textual content features used by previous works \cite{DBLP:conf/civr/YangMHYYL07}\cite{DBLP:journals/jodsn/DeldjooECGPQ16} are gradually broken through by recent findings in computer vision area \cite{DBLP:conf/eccv/PerronninSM10} \cite{DBLP:conf/iccv/WangS13a} \cite{DBLP:conf/cvpr/XuYH15}. Enlightened by this, we propose to extract diverse non-textual content features from videos in terms of audio, scenes, and motions to model the fact that users are attracted by videos in different ways \cite{DBLP:journals/tkde/HoilesAK17}. We assume the informative non-textual features useful for personalised recommendation include audio (e.g., scary videos usually have similar sound effect), scene (e.g., the interstellar movies usually have similar image background) which is encoded in images, and motion (e.g., many romantic movies have the motion of ``kiss'') which is represented as series of sequential images. In particular, the non-textual content features MFCC\cite{aly2013axes}, SIFT\cite{DBLP:journals/ijcv/RussakovskyDSKS15}\cite{chen2009mosift}, IDT\cite{DBLP:conf/iccv/WangS13a}, and CNN\cite{DBLP:journals/corr/SimonyanZ14a} that have achieved noticeable successes in video analysis tasks \cite{aly2013axes}\cite{DBLP:conf/iccv/WangS13a}\cite{DBLP:conf/cvpr/XuYH15} recently are extracted to work with the hybrid models in our work. Their individual details are described as follows.

\textbf{1. MFCC (mel-frequency cepstral coefficients)}

MFCC measure the audio changes in the sound track. We use MFCC to capture the audio contents within the videos. Our MFCC features are extracted as follows: 1) down-sampling the audio track of a video to $16$ kHz with $16$ bit resolution;  2) using a window size of $25$ ms and a step size of $10$ ms to set the signal extractor with $13$ channels; and 3) concatenating the first, second derivatives, and the energy of each signal to form a $40$-dimension vector.

\textbf{2. SIFT (scale invariant feature transform)}
	
SIFT \cite{DBLP:journals/ijcv/Lowe04} quantizes the texture information inside the images. In our study, we use two variants of SIFT to capture the scene and motion information in the videos respectively. They are OSIFT (opponent SIFT) \cite{DBLP:journals/pami/SandeGS10} and MoSIFT (motion SIFT) \cite{chen2009mosift}. OSIFT applies the light color change and shift on the original RGB color space to capture more robust scene contents in the video frames. An OSIFT feature has $384$ dimensions. MoSIFT uses the optical flow between frames to capture the motion contents from in the videos. A MoSIFT feature has $256$ dimensions.
	
\textbf{3. IDT (improved dense trajectory)}

IDT \cite{DBLP:conf/iccv/WangS13a} uses dense sampling and camera motions removing techniques to capture the motion contents in the videos. In previous studies \cite{aly2013axes}\cite{DBLP:conf/cvpr/XuYH15}, IDT performs better than MoSIFT. An IDT feature has $426$ dimensions.

\textbf{4. CNN (convolutional neutral network)}

CNN uses the deep convolutional neural network and the large-scale labeled datasets to learn how the human classifies an image. One recent research \cite{DBLP:conf/aaai/HeM16} shows that using a pre-trained CNN on ImageNet \cite{DBLP:journals/ijcv/RussakovskyDSKS15} to extract features make the product recommendation have the visually-aware ability. Inspired by this, we use the pre-trained CNN model from the VGG group\cite{DBLP:journals/corr/SimonyanZ14a} to extract tensors from the pool$_5$ layer with spatial pooling \cite{DBLP:journals/pami/HeZR015}. Accordingly, each sampled frame has $49$ CNN features with $512$ dimensions.

Unlike MFCC, MoSIFT and IDT which take the whole audio or video file as input, OSIFT and CNN are only applicable for the images.  Following\cite{aly2013axes,DBLP:conf/cvpr/XuYH15}, we fetch $5$ frames per second from the video. After all the raw features are extracted, we apply SSR (signed squared root) \cite{aly2013axes} to normalize these features for further processing.
\begin{table}[!hbt]
	\begin{center}
		\begin{tabular}{|>{\raggedright} p{40pt} |>{\centering} p{60pt} |>{\centering} p{60pt} |}
			\hline
			\textbf{Feature} & \textbf{Encoder} & \textbf{Dimension} \tabularnewline \hline
			MFCC & \multirow{5}{60pt}{\centering FV} & 10240 \tabularnewline \cline{1-1} \cline{3-3}
			OSIFT & & 98304 \tabularnewline \cline{1-1} \cline{3-3}
			MoSIFT & & 68608 \tabularnewline \cline{1-1} \cline{3-3}
			IDT & & 128304 \tabularnewline  \cline{1-1} \cline{3-3}
			\multirow{2}{40pt}{CNN}& & 131072 \tabularnewline \cline{2-3}
			& VLAD & 65536 \tabularnewline \hline
		\end{tabular}
	\end{center}
	\caption{The dimensions of the encoded non-textual content vectors.}
	\label{tbl:feconfig}
	\vspace{-15pt}
\end{table}

Each video converts into a feature tensor after the extraction no matter which non-textual content is used. Nevertheless, the feature tensor is not the ideal format for the content-side learning in hybrid models \cite{DBLP:conf/kdd/WangB11}\cite{DBLP:conf/nips/OordDS13}\cite{DBLP:conf/kdd/WangWY15}\cite{DBLP:conf/aaai/HeM16}, so these tensors are encoded to vectors for model training. In particular, we apply two state-of-the-art encoding methods, fisher vector (FV)\cite{DBLP:conf/eccv/PerronninSM10} and VLAD\cite{DBLP:journals/pami/JegouPDSPS12}, to transform feature tensors to vectors based on the reported settings \cite{aly2013axes}\cite{DBLP:conf/iccv/WangS13a}\cite{DBLP:conf/cvpr/XuYH15}. These vectors are of real values and suitable for linear models \cite{DBLP:conf/eccv/PerronninSM10,DBLP:journals/pami/JegouPDSPS12}. The encoding methods and the resulting dimensions for individual non-textual content features are recorded in Table~\ref{tbl:feconfig}. We notice that encoded non-textual content features are of high dimensions, which makes them hard work with hybrid recommender models. Thus, principle component analysis (PCA) is applied to reduce the dimensions of encoded non-textual content features to $4000$.

\section{Video Recommendation}

In this section, existing recommender models' performance in both in-matrix and out-of-matrix scenarios is first presented. Based on the performance comparison, we analyze why existing recommender models cannot deliver effective video recommendations with arbitrary content features. To address the problem, an improved recommender model, collaborative embedding regression (CER), that can work with any single content feature is proposed. Based on CER, a priority-based late fusion method (PRI) is further developed. PRI prioritizes content features by validation accuracy and assigns enough large weights to the content features of high priorities, which can further improve recommendation accuracy in our later evaluations.

\subsection{Existing Models on Arbitrary Content Features}
\begin{table}[!htb]
	\begin{center}
		\begin{tabular}{|>{\raggedright} p{100pt} |>{\raggedright} p{60pt}|}
			\hline
			\textbf{Model} & \textbf{Content Feature}\tabularnewline \hline
			WMF, BPR & N/A\tabularnewline\hline
			CDL, VBPR, CTR & WORD, META \tabularnewline\hline
			CDL, VBPR, DPM & MFCC\tabularnewline\hline
			CDL, VBPR & CNNFV\tabularnewline\hline
		\end{tabular}
	\end{center}
	\caption{Recommender models and single content features in reproduction.}
	\label{tbl:reproduction}
  \vspace{-20pt}
\end{table}
Given rich content features, an interesting question naturally arises: how do state-of-the-art recommender models perform with arbitrary content features in top-$k$ recommendations. To explore the results, we evaluate the in-matrix and out-of-matrix accuracy of representative WMF and BPR-based recommender models using the MovieLens 10M dataset\cite{DBLP:journals/tiis/HarperK16}. In this preliminary evaluation, models are not only training with their specific content features but also with other ones if possible. The details about recommender models and content features are listed in Table~\ref{tbl:reproduction}.
\begin{figure}[!hbt]
	\centering
	\includegraphics[width=7cm]{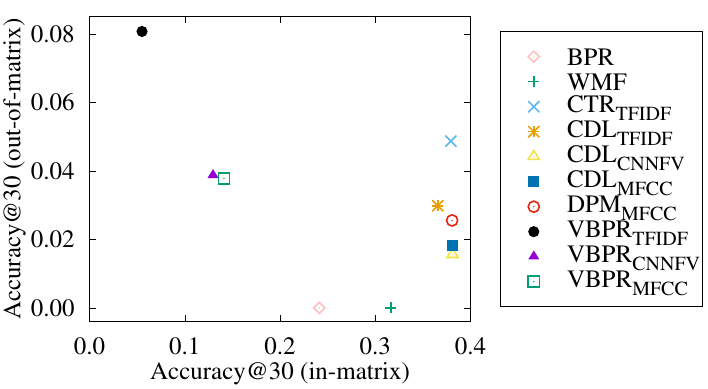}
  \vspace{-5pt}
	\small\caption{Accuracy of existing hybrid recommender models in both in-matrix and out-of-matrix scenarios, where positions near the top right corner indicate better overall performance. To clearly display positions of the models that only support in-matrix recommendation, we shift the origin of the vertical axis to a higher position.}
	\label{fig:ioas}
  \vspace{-10pt}
\end{figure}

The recommender models listed in Table~\ref{tbl:reproduction} were tested in both in-matrix and out-of-matrix settings with their optimal settings according to the reports in \cite{DBLP:conf/icdm/HuKV08,DBLP:conf/uai/RendleFGS09,DBLP:conf/kdd/WangB11,DBLP:conf/nips/OordDS13,DBLP:conf/kdd/WangWY15,DBLP:conf/aaai/HeM16}. We use in-matrix and out-of-matrix accuracy as the coordinates to draw combinations of models and content features in Fig.~\ref{fig:ioas}. Specifically, we use the evaluation metric Accuracy@$30$ to plot the points. Additionally, in the legend of Fig. \ref{fig:ioas}, the subscripts of models denote the in-use content features. Fig. \ref{fig:ioas} provides the following observations:
\begin{enumerate}
  \item \textbf{WMF-based recommender models achieved the best performance in in-matrix scenario}. In Fig.~\ref{fig:ioas}, all the WMF-based models (i.e. WMF, CTR, DPM and CDL) are located to the right of the BPR-based models (i.e. BPR and VBPR). In addition, the in-matrix performance of WMF-based models (e.g., CDL) do not vary obviously with respect to different content features. All these facts indicate that WMF-based models are better than BPR-based models for video recommendation in in-matrix scenario.
      
  \item \textbf{BPR-based recommender models achieved the best performance in out-of-matrix scenario}. Also in Fig.~\ref{fig:ioas}, given a particular content feature, the position of VBPR is always higher than that of all the WMF-based models. This shows that VBPR is currently the most effective model in out-of-matrix scenario, yet indicates the content-side learning inside existing WMF-based models are not effective for out-of-matrix video recommendation.
\end{enumerate}

Our reproduction experiment shows that none of the existing recommender models achieve superior accuracy in both in-matrix and out-of-matrix scenarios even with their nominated content features. In the next section, we will propose a general recommender model, collaborative embedding regression (CER), to address the issue.

\subsection{Collaborative Embedding Regression}

\begin{table}[!htb]
  \begin{center}
    \begin{tabular}{|>{\centering} p{40pt} |>{\raggedright} p{140pt}|} \hline
      \textbf{Denotation} & \centering \textbf{Explanation} \tabularnewline \hline
      $u_i$ & $i^{th}$ user's latent vector \tabularnewline \hline
      $h_j$ & $j^{th}$ video's latent vector \tabularnewline \hline
      $f_j$ & $j^{th}$ video's content vector \tabularnewline \hline
      $h_j'$ & $j^{th}$ video's content latent vector \tabularnewline \hline
      $\epsilon{}_{j}$ & the offset vector between $j^{th}$ video's latent vector and content latent vector \tabularnewline \hline
      $r_{ij}$ & $i^{th}$ user's rating on $j^{th}$ video \tabularnewline \hline
      $c_{ij}$ & $i^{th}$ user's confidence on $j^{th}$ video \tabularnewline \hline
      $I_k$ & identify matrix with $k$ dimensions \tabularnewline \hline
      $I_k^{-1}$ & the inverse of matrix $I_k$ \tabularnewline \hline
      $E$ & matrix that embeds $f_j$ to $h_j'$ \tabularnewline \hline
      $\lambda{}$ & the hyper parameter for regularization  \tabularnewline \hline
      $\pi{}_l$ & the weight of $l^{th}$ content feature in the late fusion\tabularnewline \hline
      $||\cdot{}||_{F}$ & Frobenius norm \tabularnewline \hline
    \end{tabular}
  \end{center}
  \caption{Denotations of the variables used in collaborative embedding regression (CER).}
  \vspace{-20pt}
\end{table}

Existing WMF-based models\cite{DBLP:conf/kdd/WangB11,DBLP:conf/nips/OordDS13,DBLP:conf/kdd/WangWY15} are designed for document or music recommendation tasks where single content features are often strong enough to support content-side learning. In particular, CTR\cite{DBLP:conf/kdd/WangB11} incorporates the word vectors with WMF and performs content-side learning by latent Dirichlet allocation (LDA). Since the optimization of LDA is based on word count, CTR naturally fails to support non-textual content features that are of real values. Compared to CTR, DPM\cite{DBLP:conf/nips/OordDS13} and CDL\cite{DBLP:conf/kdd/WangWY15} perform content-side learning with MFCC vectors and word vectors respectively by multiple layer perception (MLP) and stacked de-nosing auto-encoder (SDAE). However, their content-side learning aims at improving the in-matrix recommendation accuracy when ratings are sparse and implicitly requires the input feature vectors to be of non-negative values. As a result, DPM and CDL's out-of-matrix performance is poor according to Fig.~\ref{fig:ioas} even with their specific content features. Compared to documents or music, videos are associated with multiple content features that none of them can solely support effective recommendation. In this situation, a general rather than specific content-side learning is necessary for hybrid models. VBPR's performance in out-of-matrix shows that linearly embedding content features in the content-side learning is general and powerful. Based on above analysis, we propose a general WMF-based recommender model, collaborative embedding regression (CER), to achieve state-of-the-art in-matrix and out-of-matrix recommendation accuracy with arbitrary single content features.

Let $d$ denote the dimension of the content vector and $k$ denote the dimension of the latent vector. The whole generation process of CER with an individual content feature is described below.
\begin{enumerate}
  \item For each user $i$, draw a user latent vector $w_i\in{}\mathcal{R}^{k\times{}1}$:\\
   \begin{equation}\label{l:1}
   \small
     w_i\sim\mathcal{N}(0,\lambda{}_u^{-1}I_k).
   \end{equation}
  \item Generate an embedding matrix $E\sim\mathcal{N}(0,\lambda_e^{-1}I_k)$.
  \item For each video $j$:
  \begin{enumerate}
    \item Generate a content latent vector $h'_j\in\mathcal{R}^{k\times{}1}$:
     \begin{equation}\label{l:2}
   \small
     h'_j=E^Tf_j.
   \end{equation}
    \item Draw a latent video offset vector $\epsilon_j\sim{}\mathcal{N}(0,\lambda_v^{-1}I)$, and then set the video latent vector as:
     \begin{equation}\label{l:3}
   \small
    h_j=h'_j+\epsilon_j.
   \end{equation}
 \end{enumerate}
   \item  For each user-video pair $(i,j)$, draw the rating:
 \begin{equation}\label{l:4}
   \small
   r_{ij}\sim \mathcal{N}(w_i^Th_j,c_{ij}^{-1}).
   \end{equation}
\end{enumerate}
where $I_k$ is an identity matrix, $f_j\in{}\mathcal{R}^{d\times{}1}$ is a feature vector, $E\in{}\mathcal{R}^{d\times{}k}$ is an embedding matrix, and $c_{ij}$ is the confidence parameter for the user-item pair $(i,j)$. Following\cite{DBLP:conf/kdd/WangB11,DBLP:conf/kdd/WangWY15}, the value of $c_{ij}$ is defined below:
\begin{equation}
c_{ij} =
\begin{cases}
1, & \text{if $r_{ij}=1$}\\
0.01, & \text{if $r_{ij}=0$}
\end{cases}
\end{equation}

Note that, in step $3(a)$, we use linear embedding instead of non-linear learning adopted by CTR, DPM and CDL. This is more general for content-side learning with arbitrary content features from videos. In step $3(b)$, $h'_j$ that embeds content features to latent vectors serves as the bridge between content-side and collaborative-side learning.

\textbf{Learning the parameters}. The lowest mean absolute error (MAE) of rating prediction is achieved when the jointly posterior probability of $W$, $H$ and $E$ is maximized. However, directly computing the full posterior of parameters is intractable. As a result, CER is trained by minimizing the negative log-likelihood in this paper as follows:
\begin{equation}
\begin{aligned}
&\sum\limits_{i=1}^m\sum\limits_{j=1}^n\frac{c_{ij}}{2}(w_i^Th_j-r_{ij})^2+\frac{\lambda{}_u}{2}\sum\limits_{i=1}^mw_i^Tw_i+\\
&\frac{\lambda{}_v}{2}\sum\limits_{j=1}^n(h_j-E^Tf_j)^T(h_j-E^Tf_j)+\frac{\lambda{}_e}{2}||E||_F^2,
\end{aligned}
\label{eq:objf}
\end{equation}

where $\lambda_u$, $\lambda_v$ and $\lambda_e$ are the hyper parameters and $||\cdot{}||_F$ denotes the Frobenius norm. When hyper parameters are given, the optimal latent vectors $w_i$ and $h_j$ as well as the embedding matrix $E$ are learned by performing the alternating least squares (ALS). Specifically, given the current estimation of $E$, we calculate the derivatives with respect to $w_i$ and $h_j$, set them to zeros, and apply following updates for $w_i$ and $h_j$ in each iteration:
\begin{equation}
\begin{aligned}
w_i&\leftarrow{}(HC_iH^T+\lambda{}_uI_k)^{-1}HC_iR_i\\
h_j&\leftarrow{}(WC_jW^T+\lambda{}_vI_k)^{-1}(WC_jR_j+\lambda{}_vE^Tf_j)\\
\end{aligned}
\end{equation}

where $W=(w_i)^m_{i=1}\in{}\mathcal{R}^{k\times{}m}$ is the matrix concatenated by user latent vectors, $H=(h_j)^n_{j=1}\in{}\mathcal{R}^{k\times{}n}$ is the matrix concatenated by video latent vectors, and $F=(f_j)^n_{j=1}\in{}\mathcal{R}^{d\times{}n}$ is the content feature matrix. For user $i$, $C_i\in{}\mathcal{R}^{n\times{}n}$ is a diagonal matrix with $c_{ij}$, $j=1\cdots{},n$ as the diagonal elements, $R_i\in{}\{0,1\}^{n\times{}1}$ is a vector with $r_{ij}$, $j=1\cdots{},n$ as its elements. For video $j$, $C_j$ and $R_j$ are similarly defined.

After $W$ and $H$ are updated, the derivatives with respect to $E$ are computed and set to zero. Then, $E$ is applied the following update:
\begin{equation}
\begin{aligned}
E&\leftarrow{}(\lambda_vFF^T+\lambda_eI_d)^{-1}(\lambda_vFH^T).
\end{aligned}
\end{equation}

Similar to CTR and CDL, CER supports both in-matrix and out-of-matrix rating prediction. For in-matrix predictions, given a user-video pair $(i,j)$, the rating $\hat{r}_{ij}$ is estimated as $w^T_i(E^Tf_j+\epsilon_j)$. For out-of-matrix prediction, the rating $\hat{r}_{ij}$ is predicted as $w^T_iE^Tf_j$ since no offset is observed. In summary, CER's rating predictor is defined as:
\begin{equation}
\hat{r}_{ij} =
\begin{cases}
w_i^Th_j, & \text{in-matrix setting}\\
w_i^TE^Tf_j, & \text{out-of-matrix setting}
\end{cases}
\end{equation}

\subsection{Multiple Feature Fusion}
\label{sec:fusion}
CER model is designed to work with single content features as existing hybrid models \cite{DBLP:conf/kdd/WangB11}\cite{DBLP:conf/nips/OordDS13}\cite{DBLP:conf/kdd/WangWY15}\cite{DBLP:conf/aaai/HeM16}. Based on CER, we further explore how to use multiple content features to improve the video recommendation accuracy. Specially, we discuss three feature fusion methods that possibly facilitate the video recommendation with multiple content features.

The first method concatenates multiple content feature vectors associated with same videos into single big vectors and then feeds the big vectors into CER for model training. Assuming there are $L$ content features in total, the concatenation is denoted as follow:
\begin{equation}
f_j\leftarrow{}[f_j^1, f_j^2, \dots{}, f_j^L].
\end{equation}

This fusion method is expected to learn the shared latent factors among the concatenated features. It does not introduce any modification on the objective function of CER, but it will significantly increase the training cost because the time complexity of CER's training is proportional to the dimension of the feature vector $f_j$.

The second method adds multiple content latent vectors $h'^{l}_j$ together, as done in CKE\cite{DBLP:conf/kdd/ZhangYLXM16}. The content latent vectors in the generation process of CER are thus redefined as:
\begin{equation}
h'_j=\sum\limits_{l=1}^L h'^{l}_j =\sum\limits_{l=1}^LE^lf_j^l.
\end{equation}
Compared to the first method, the second method compresses the dimension so that the training is faster, but it needs to modify the objective function of the CER by adding regularization terms of all the embedding matrices. In addition, updating formulas of model parameters need to change accordingly.

The first two methods are early fusion. They aim at mapping multiple content feature spaces to a shared homogeneous one. However, the textual, audio, visual and motion information contained in videos are quite diverse and heterogeneous. As shown in Fig.~\ref{fig:ioas}, the out-of-matrix accuracy of the best model varies greatly with respect to the different content features. This indicates that the construction of a shared latent space without losing some important and meaningful content patterns is usually unreasonable. Moreover, early fusion methods require re-training models when the deployed content features are changed (e.g., adding new ones). Based on above discussions, we think early fusion methods tend to be poor for video recommendation with multiple content features.

In a recent finding from other areas \cite{aly2013axes}\cite{DBLP:conf/cvpr/XuYH15}, the performance divergence caused by heterogeneous content features can be addressed by late fusion, which fuses prediction scores from multiple content features to form more relevant ones. Inspired by this finding, we think late fusion has the potential to improve the video recommendation with multiple content features. We thus propose the third method to calculate fusion scores for videos. The score calculation is as follows:
\begin{equation}
\small
\bar{r}_{ij}=\sum\limits_{l=1}^L \pi_l\hat{r}_{ij}^l,
\end{equation}
where $L$ is the number of content features; $\pi_l$ is the weight of the $l^{th}$ content feature; $\hat{r}_{ij}^l$ is the predicted rating based on the $l^{th}$ content feature. How to compute the weights is the major challenge of late fusion. A naive solution to compute weights, namely average fusion, is to treat each content feature equally. Recall that Fig.~\ref{fig:ioas} shows significant performance divergences between different content features. Average fusion tends to neglects the divergences so as to lead to inferior fusion performance. Another solution is to learn the weights using a learning-to-rank method \cite{DBLP:conf/icml/CaoQLTL07}. However, as shown in our later experiments, learning-to-rank models cannot correctly quantize the influence of each content feature in the out-of-matrix recommendation.

In this paper, we propose a priority-based late fusion method (PRI) which aims at making the weights reflect higher influences from more effective content features. In details, PRI prioritizes content features based on the validation out-of-matrix performances and generates a feature ranking list from high to low, then iteratively assigns the weight of value $\pi_l=p(1-p)^{l-1}$ to $l^{th}$ content feature in the ranking list where $p\in{}[0.5,1)$ is a hyper parameter. Note that, for any ranking position $t$ (i.e., $\forall t>0$), the inequality $\sum_{l=t+1}^L \pi_l\leqslant{}\pi_t$ holds in the related weights, which is not guaranteed in existing late fusion methods. This inequality ensures that the $l^{th}$ content feature always has higher weight than the total weight of the remaining less powerful content features. In other words, PRI allows more effective content features have more impacts in the fusion, which will be examined in our multiple feature fusion evaluation.

To clearly illustrate the computation of PRI's weights, we present an example in  Table~\ref{tbl:fusion_example} where four content features are given and ranked. Note that, hybrid models (including our CER) with different content features achieve the similar recommendation performances in in-matrix scenario due to the absolute dominance of CF \cite{DBLP:conf/kdd/WangB11}, so we only apply PRI in out-of-matrix scenario with multiple content features.
\begin{table}[!hbt]
	\begin{center}
		\begin{tabular}{|>{\raggedright} p{32pt} |>{\raggedright} p{32pt} |>{\centering} p{32pt} |>{\centering} p{32pt} |>{\centering} p{32pt} |}
			\hline
			Feature & META & CNNFV & IDT & MFCC \tabularnewline \hline
			$l$    & 1 & 2 & 3 & 4 \tabularnewline\hline
			$\pi_l$  & 0.5 & 0.25 & 0.125 & 0.0625 \tabularnewline\hline
		\end{tabular}
	\end{center}
	\caption{An example of the weights used by PRI when $p$ is set to $0.5$.}
	\label{tbl:fusion_example}
	\vspace{-20pt}
\end{table}

\section{Experiments}

In this section, we first introduce the datasets, the comparison models and methods, and the experimental settings in our evaluations. Then, we report the experimental results in terms of single content recommending and multiple content fusion. In addition, we present some important findings from our experiment that are emphasized by the bold font.

\subsection{Dataset Description}
We used the MovieLens 10M \cite{DBLP:journals/tiis/HarperK16} and the Netflix 100M \footnote{\url{https://en.wikipedia.org/wiki/Netflix_Prize}} as the base datasets for our empirical studies. We do not use the YouTube 8M \footnote{\url{https://research.google.com/youtube8m/index.html}} and the YouTube faces DB \footnote{\url{https://www.cs.tau.ac.il/~wolf/ytfaces}} because they do not provide user-video interactions. Both of MovieLens and Netflix do not contain videos or links for downloading, so we attempted to collect the videos from YouTube by ourselves. After several attempts, we finally downloaded the trailers instead of the full-length videos, because most of the full-length videos are not available for downloading due to copyright restrictions. We also manually checked whether the trailers were really for the original full-length videos. For those trailers that are mismatched, we used the available video clips on YouTube from the original videos instead. By these means, we collected $10380$ videos of the $10682$ movies for the MovieLens 10M dataset, and $4831$ videos of the $17770$ movies for the Netflix 100M. The fps of videos is $24$. The minimum, maximum and mean lengths of videos in for different datasets are listed in Table \ref{tbl:video}. Videos' widths are resized to $240$ pixels for accelerating the non-textual content feature extraction. Their heights were resized proportionally according to the aspect ratio.
\begin{table}[!htb]
  \begin{center}
    \begin{tabular}{|>{\raggedright} p{40pt} |>{\centering} p{60pt} |>{\centering} p{60pt} |}\hline
      & MovieLens & Netflix \tabularnewline \hline
      length$_{min}$ & 14.58s & 14.90s \tabularnewline \hline
      length$_{max}$ & 578.97s & 370.73s \tabularnewline \hline
      length$_{mean}$ & 127.59s & 127.16s \tabularnewline \hline
    \end{tabular}
  \end{center}
  \caption{The statistics of video lengths on both datasets}
  \label{tbl:video}
  \vspace{-20pt}
\end{table}

For the MovieLens dataset, we use the movie IDs from IMDB\footnote{\url{http://www.imdb.com}} that the MovieLens dataset provides to collect the textual content of the videos. Specifically, we crawled the movie plots, actors, directors, companies, languages and genres for the collected videos. After data collecting and natural language processing, we firstly created a corpus where each document is made up of the corresponding movie title and plot. After that, the top $20000$ words in the corpus were selected as the vocabulary according to the global TF-IDF values, following \cite{DBLP:conf/kdd/WangB11} and \cite{DBLP:conf/kdd/WangWY15}. Given the vocabulary, a word vector for each movie was generated by counting the word frequencies. The remaining textual data include actors, directors, languages, companies, genres and other meta items. They form a meta vector that is binary for each movie. We wanted to compare the effectiveness of the textual features fairly, so we again selected top $20000$ meta items to generate the meta vectors. The Netflix dataset has only the titles for the movies. We used these titles to crawled the same types of textual data from IMDB as done for the MovieLens dataset, and apply the same text feature extraction process on the collected data. 

Given the collected video list, the ratings associated with the missing videos and the users that have no ratings for the collected videos were removed. We transformed the remaining ratings in both datasets into $\{0,1\}$ to model the implicit feedback as \cite{DBLP:conf/kdd/WangB11} \cite{DBLP:conf/kdd/WangWY15} \cite{DBLP:journals/tkde/ChenW0Y16}. Specifically, we changed the ratings with the highest value $5$ to $1$ and all the other ratings to $0$. After that, we got a MovieLens implicit dataset that has $1,543,593$ positive feedbacks and a Netflix implicit feedback dataset that has $13,016,825$ positive feedbacks. In these two implicit feedback datasets, the positive feedbacks fill $0.21\%$ and $0.56\%$ of the implicit rating matrix respectively. To clearly show the changes between original and processed data, we summarize the important changes between them in Table \ref{tbl:dataset}. The comparison between the rating density and the positive density shows that the implicit feedback transformation makes the training positives even less, which has been widely observed in the past literatures \cite{DBLP:conf/kdd/WangB11}\cite{DBLP:conf/kdd/WangWY15}\cite{DBLP:conf/aaai/HeM16}. We made the experimental data and code publicly available for reproducible purpose\footnote{\url{https://github.com/domainxz/top-k-rec}}.
\begin{table}[!htb]
    \begin{center}
    \begin{tabular}{|>{\raggedright} p{30pt} |>{\raggedleft} p{40pt} |>{\raggedleft} p{40pt} |>{\raggedleft} p{40pt} |>{\raggedleft} p{40pt} |}\hline
        \multirow{2}{*}{Index} & \multicolumn{2}{c|}{MovieLens} & \multicolumn{2}{c|}{Netflix} \tabularnewline \cline{2-5}
        & original & processed & original & processed \tabularnewline \hline
        \#user   & $69,878$ & $69,878$ & $480,189$ & $478,624$ \tabularnewline \hline
        \#video  & $10,682$ & $10,380$ & $17,770$ & $4,831$ \tabularnewline \hline
        \#rating & $10,000,054$ & $9,988,676$ & $100,480,507$ & $62,714,775$ \tabularnewline \hline
        rating density & $1.34\%$ & $1.38\%$ & $1.18\%$ & $2.71\%$ \tabularnewline \hline
        \#implicit positive & $1,544,812$ & $1,543,593$ & $23,168,232$ & $13,016,825$ \tabularnewline \hline
        positive density & $0.21\%$ & $0.21\%$ & $0.27\%$ & $0.56\%$ \tabularnewline \hline
    \end{tabular}
    \end{center}
    \caption{The statistics of the MovieLens and the Netflix datasets in evaluation}
    \label{tbl:dataset}
    \vspace{-20pt}
\end{table}

\subsection{Experimental Setup}
\subsubsection{Evaluation Metric}
We adopt the evaluation metric Accuracy@$k$ used in \cite{DBLP:conf/recsys/CremonesiKT10,DBLP:journals/pvldb/YinCLYC12,DBLP:conf/icde/WangYSCXZ16,DBLP:conf/sigir/WangYHWDN18} to evaluate the top-$k$ video recommendation accuracy in our experiment. The Accuracy@$k$ metric reveals the ratio between the number of the overall positives that are correctly predicted by the personalized recommendations and the number of the total ground truth positives, where a higher value means better performance. In previous works \cite{DBLP:conf/kdd/WangB11} and \cite{DBLP:conf/kdd/WangWY15}, the value of $k$ was selected from $\{50,100,150,200,250,300\}$, which was too large for a user to receive at once in a real-world video recommender system \cite{gomez2015netflix}. Therefore, $k$ is selected from $\{5,10,15,20,25,30\}$ in our evaluation. Given the value of $k$, the steps to calculate Accuracy@$k$ value are as follows:
\begin{itemize}
  \item [1.] Inferring the users' ratings on their unrated videos by the latent vectors and the content vectors;
  \item [2.] Generating a ranking list of the unrated videos for each user according to the descending order of the corresponding inferred ratings; 
  \item [3.] Selecting the top-$k$ videos from the ranking list to form the recommendation for each user;
  \item [4.] Counting the number of hits on the test set $D_{test}$: for each user-video pair $(i,j)$ in the test set $D_{test}$, if video $j$ appears in user $i$'s top-k recommendation list, we have a hit (i.e., the ground truth video is recommended to the user), otherwise we have a miss;
  \item [5.] Calculating the overall Accuracy@$k$ by Eq. \eqref{eq:acc@k}.
  \begin{equation}
  \label{eq:acc@k}
  Accuracy@k=\frac{\#Hit@k}{|D_{test}|}
  \end{equation}
  where $\#Hit@k$ denotes the total number of hits in the test set, and $|D_{test}|$ is the number of all test cases.
\end{itemize}

We adopt 5-fold cross validation method. Thus, we report the mean Accuracy@$k$ of each recommender model or method.

\subsubsection{Data Preparation}
Each dataset in our evaluation is divided into the training set, the in-matrix test set, and the out-of-matrix test set. We do the dataset division five times to measure the mean accuracy of each recommender model or method in our evaluation. We apply the following steps to make these sets occupying $60\%$, $20\%$, $20\%$ of the total positive ratings respectively:
\begin{itemize}
  \item[(a)] Splitting the unique video ids in the dataset into five folds randomly and uniformly, then splitting the ratings into five folds according to the split video ids;
  \item[(b)] Selecting the rating fold from (a) one by one as the out-of-matrix test set;
  \item[(c)] Merging the rest four rating folds from (b), then randomly re-splitting them into four folds uniformly;
  \item[(d)] Selecting one rating fold from (c) as the in-matrix test set;
  \item[(e)] Merging the rest three folds from (d) as the training set for model training.
\end{itemize}

In order to tune the hyper parameters of each model, the training set was further divided: $90\%$ of the training ratings were used to train the model and $10\%$ of the training ratings were used for validation. The in-matrix test set shares the same video ids with the training set. It was used to evaluate the accuracy of collaborative filtering (CF) side in each model. The out-of-matrix test set has totally different video ids from the training set's. It was used to evaluate the effectiveness of each content feature and the accuracy of content side in each hybrid recommender model.

Based on the video titles we have, we construct the sparse text vectors (denoted as \textbf{ST}) for the videos to simulate the situation where limited texts are available such as the users' generated or uploaded videos.. The accuracy measured on ST will be used as the baseline to show how the non-textual contents improve the recommendation accuracy when the recommender system lacks enough textual contents.

We also design two rich content feature sets to work with the fusion methods. These rich content feature sets aim to explore the impact on accuracy when the textual contents are absent. The settings of these rich content feature sets are as follows:
\begin{itemize}
  \item \textbf{Rich content feature set} contains all the content features we extract in this study. This set is used to examine whether a fusion method can fuse textual contents and non-textual contents properly. 
  \item \textbf{Rich non-textual content feature set} contains all the non-textual content features we extract in this study. This set is used to examine whether a fusion method can use the non-textual contents to achieve the same performance as the textual contents. 
\end{itemize}

\subsubsection{Comparison Model and Method}

We compare our proposed CER model with the six state-of-the-art recommender modelds. For fair comparison, the dimension of the latent vectors in all the models is set to $50$. Below, we provide the details of the evaluated models as well as their hyper parameter settings in our experiment for the reproducible purpose.

\textbf{Weighted Matrix Factorization (WMF)}\cite{DBLP:journals/computer/KorenBV09} is a classical collaborative filtering (CF) model that works in in-matrix setting. In our experiment, WMF achieved its highest accuracy with $\lambda_u=0.01, \lambda_v=0.01$.

\textbf{Collaborative Topic Regression (CTR)}\cite{DBLP:conf/kdd/WangB11} combines CF with textual content. CTR learns the content latent vectors from word vectors using LDA. In our experiment, CTR was trained with both word and meta vectors, and it achieved the highest accuracy with $\lambda{}_u=0.1$, $\lambda{}_v=10$.

\textbf{DeepMusic (DPM)}\cite{DBLP:conf/nips/OordDS13} originally combines CF with audio content. DPM uses MLP to learn the content latent vectors from the MFCC vectors. In our experiment, DPM was extended to work with all the content vectors in the rich set, and it achieved the highest accuracy with $\lambda_u=0.1$ and $\lambda_v=10$.
	
\textbf{Collaborative Deep Learning (CDL)}\cite{DBLP:conf/kdd/WangWY15} originally combines CF with textual content. CDL uses stack denoising auto-encoder (SDAE) to learn the content latent vectors from the word vectors. In fact, SDAE can process the non-textual contents by replacing the binary visible layer with Gaussian visible layer. We therefore modified CDL to work with both textual and non-textual content vectors in our experiment. Referring to \cite{DBLP:conf/kdd/WangWY15}, we chose the three-layer architecture for content side of CDL. The chosen architecture achieved its highest accuracy with $\lambda{}_u=0.1$, $\lambda{}_v=10$ and $\lambda{}_n=1000$.
	
\textbf{Bayesian Personalized Ranking (BPR)}\cite{DBLP:conf/uai/RendleFGS09} uses pair-wise optimization to perform CF. Similar to WMF, BPR is only applicable in the in-matrix scenario. In our experiment, BPR achieved its highest accuracy with $\lambda_u=0.0025$, $\lambda{}_i=0.0025$, $\lambda_j=0.00025$ and $\lambda_b=0.0$.
	
\textbf{Visual Bayesian Personalized Ranking (VBPR)}\cite{DBLP:conf/aaai/HeM16} is an extension of BPR to combine CF with scene content. VBPR can work with any individual content feature. VBPR achieved its highest accuracy with $\lambda{}_u=0.0025$, $\lambda{}_p=0.0025$, $\lambda{}_i=0.0025$, $\lambda{}_j=0.00025$, $\lambda{}_b=0.0$ and $\lambda{}_e=0.0$.

\textbf{Collaborative embedding regression (CER)} is proposed in this paper to combine CF with any individual content feature. CER uses linear embedding to learn the content latent vectors from the content vectors. CER achieved its highest accuracy with $\lambda{}_u=0.1$, $\lambda{}_v=10$ and $\lambda{}_e=1000$.

We also compare the proposed priority-based late fusion method with six widely used fusion methods. The descriptions of these fusion methods are as follows:

\textbf{Accuracy fusion (ACC)} is a late fusion method that uses the validation accuracy of the individual content features as the weights to calculate the fusion rating for the videos. We use ACC as a heuristic baseline to validate whether the exponential weights in our priority-based fusion method are effective.

\textbf{Average fusion (AVG)} is a late fusion method that averages the predicted ratings from different content features.

\textbf{Ranking SVM (SVM)}\cite{DBLP:journals/ftir/Liu09} is a late fusion method that applies the point-wise learning-to-rank process.

\textbf{Ranking BPR (BPR)}\cite{DBLP:conf/uai/RendleFGS09} is a late fusion method that applies the pair-wise learning-to-rank process.

\textbf{Early fusion by content vector concatenating (ECT)} that has been applied in \cite{DBLP:conf/aaai/HeM16} is an early fusion method presented in Section~\ref{sec:fusion}. It concatenates all the feature vectors to learn the unified latent space by linear embedding.

\textbf{Early fusion by latent content vector stacking (ESK)} that has been applied in \cite{DBLP:conf/kdd/ZhangYLXM16} is an early fusion method presented in Section~\ref{sec:fusion}. It adds all the content latent vectors in a element-wise way to learn the unified latent space.

\textbf{Priority-based fusion (PRI)} is the late fusion method proposed in this paper. Like ACC, PRI obtains the accuracy of each content feature on the validation set first. Then, it prioritizes the content features and assign them the exponential weights like Table \ref{tbl:fusion_example}. We applied grid search on $p$ to select its best value from $\{0.5,0.6,0.7,0.8,0.9\}$. On the validation set, we found PRI achieved its best performance on both rich content feature sets with $p=0.5$.

\subsection{Evaluation on Individual Features}
We evaluate the performances of recommender models on single content features in this study. Specifically, we compare CER with state-of-the-art recommender models in both in-matrix and out-of-matrix scenarios. The comparison validates that CER is a general and effective hybrid model to work with different content features.
\subsubsection{In-matrix Evaluation}
\begin{figure}[!hbt]
	\centering
  \subfigure[MovieLens]{\includegraphics[width=4cm]{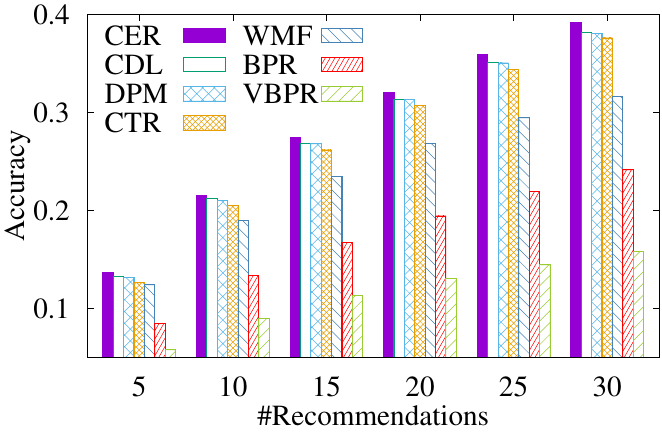}\label{fig:immethodml}}
  \subfigure[Netflix]{\includegraphics[width=4cm]{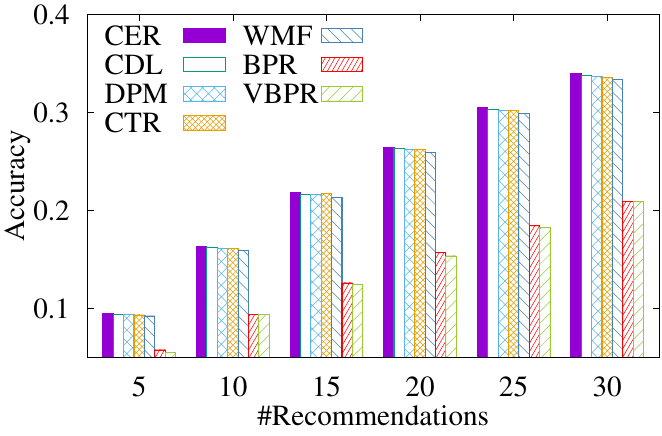}\label{fig:immethodnf}}
  \vspace{-10pt}
	\caption{In-matrix accuracy comparison between different recommender models}
	\label{fig:immethod}
  \vspace{-10pt}
\end{figure}

In the in-matrix evaluation, we study two problems as follows:
\begin{itemize}
  \item How CER performs against hybrid models with single content features in in-matrix scenario?
  \item Whether CER can generate accurate recommendations with arbitrary single content features in in-matrix scenario?
\end{itemize}
We firstly compare CER with the other recommender models that can generate in-matrix recommendations. The accuracy results on MovieLens and Netflix are plotted in Fig. \ref{fig:immethodml} and \ref{fig:immethodnf} respectively.

In Fig. \ref{fig:immethodml} and \ref{fig:immethodnf}, we only present each hybrid recommender model's highest accuracy with single content features. We observe that CER has consistently achieved higher accuracy than the other models on both datasets, which shows that \textbf{CER is capable of generating accurate in-matrix recommendation as other hybrid models}. We also observe that the accuracy gaps between the BPR-based models and WMF-based models are significant, which again validates that the WMF-based models are more effective than the BPR-based models for video recommendation in the in-matrix scenario. Additionally, we notice the accuracy gaps between pure WMF and its variants (CTR, DPM, CDL, CER) are considerable. This observation indicates that the content-side is beneficial to the collaborative-side inside a hybrid model. The comparison also shows that different content-side learning methods inside the WMF-based models have marginal impacts on in-matrix accuracy.

\begin{figure}[!hbt]
  \centering
  \subfigure[MovieLens]{\includegraphics[width=4cm]{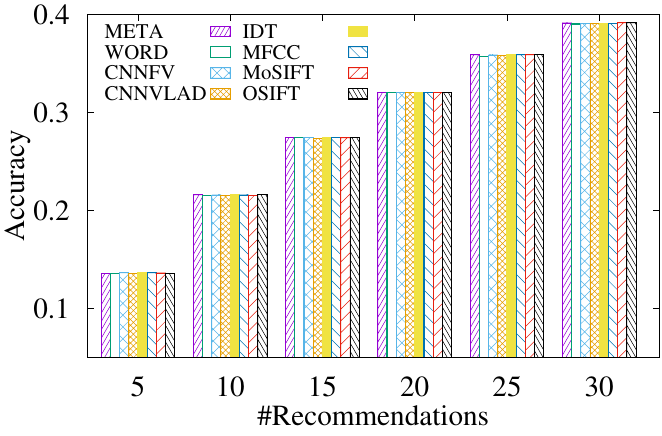}\label{fig:imcontentml}}
  \subfigure[Netflix]{\includegraphics[width=4cm]{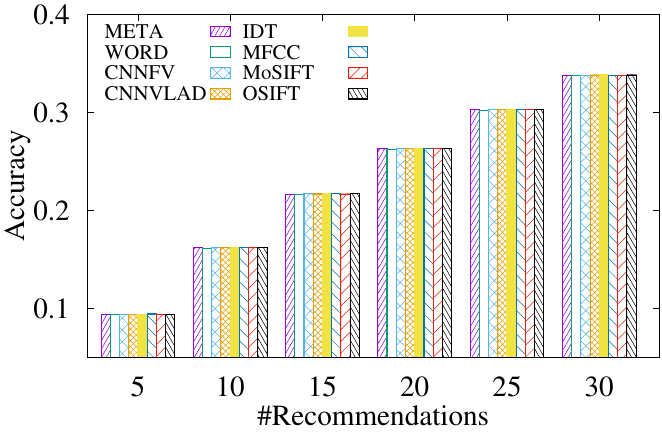}\label{fig:imcontentnf}}
  \vspace{-10pt}
  \caption{In-matrix accuracy comparison among different features}
  \label{fig:imcontent}
  \vspace{-10pt}
\end{figure}

We secondly explore where CER is able to generate accurate in-matrix recommendations with arbitrary single content features. The performances of CER with different content features are depicted in Fig. \ref{fig:imcontentml} and \ref{fig:imcontentnf}. 

From Fig. \ref{fig:imcontentml} and \ref{fig:imcontentnf}, we observe that different content features have similar individual impacts to in-matrix accuracy except WORD vectors. The results show that CER can generate accurate in-matrix recommendations with arbitrary single content features. Considering WORD vectors have been widely used by existing works \cite{DBLP:conf/kdd/WangB11}\cite{DBLP:conf/kdd/WangWY15}, the results further show that \textbf{there is almost no performance drop of changing content features inside CER in in-matrix scenario}.

\subsubsection{Out-of-matrix Evaluation}
\begin{figure*}[!t]
  \centering
  \subfigure[META]{\label{fig:ommetaml}\includegraphics[width=4cm]{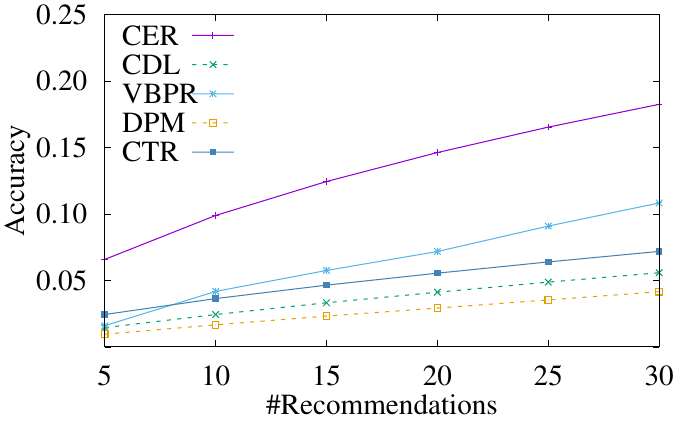}}
  \subfigure[WORD]{\label{fig:omwordml}\includegraphics[width=4cm]{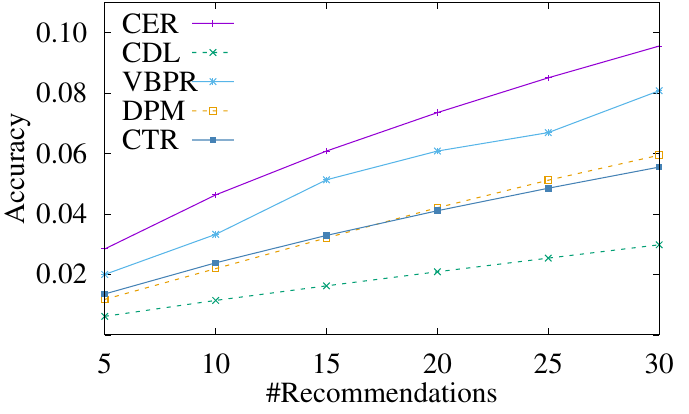}}
  \subfigure[CNNFV]{\label{fig:omcnnfvml}\includegraphics[width=4cm]{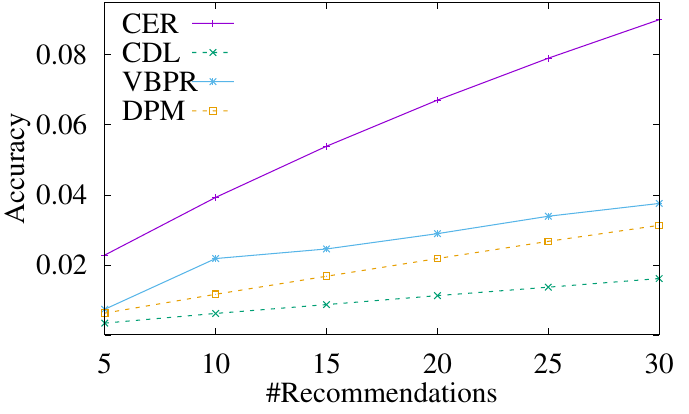}}
  \subfigure[CNNVLAD]{\label{fig:omcnnvladml}\includegraphics[width=4cm]{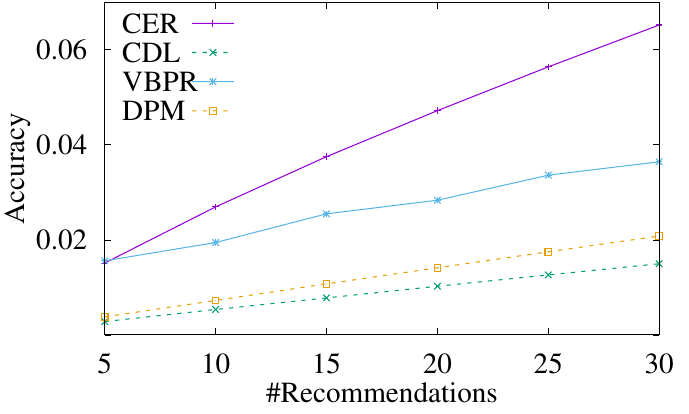}}
  \subfigure[IDT]{\label{fig:omidtml}\includegraphics[width=4cm]{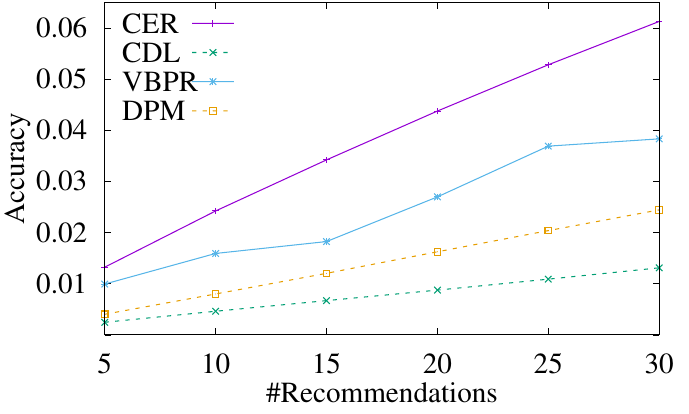}}
  \subfigure[MFCC]{\label{fig:ommfccml}\includegraphics[width=4cm]{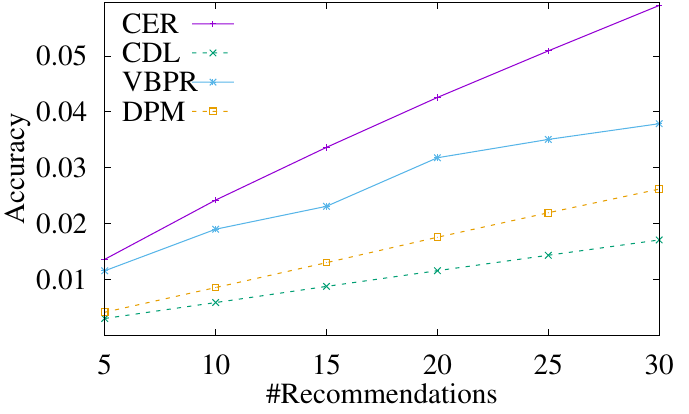}}
  \subfigure[OSIFT]{\label{fig:omosiftml}\includegraphics[width=4cm]{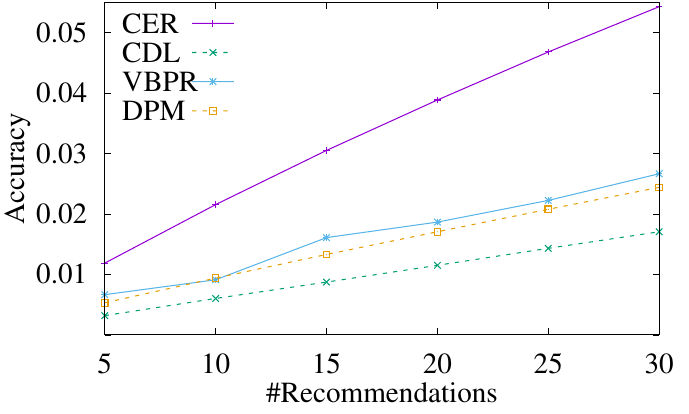}}
  \subfigure[MoSIFT]{\label{fig:ommosiftml}\includegraphics[width=4cm]{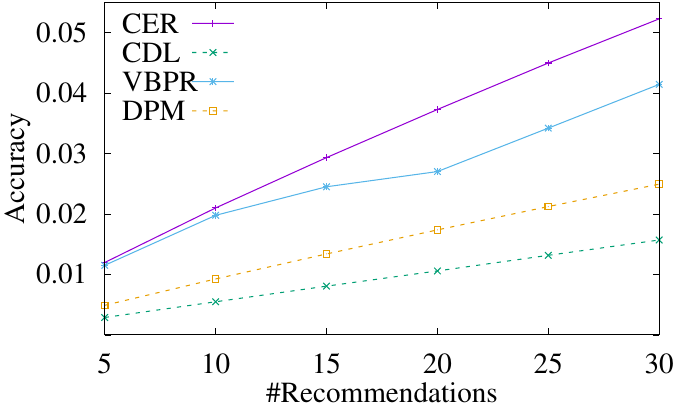}}
  \vspace{-10pt}
  \caption{Accuracy@k of different recommender models and features in out-of-matrix setting on MovieLens implicit dataset.}
  \label{fig:omaccml}
  \vspace{-10pt}
\end{figure*}

\begin{figure*}[!t]
  \centering
  \subfigure[META]{\label{fig:ommetanf}\includegraphics[width=4cm]{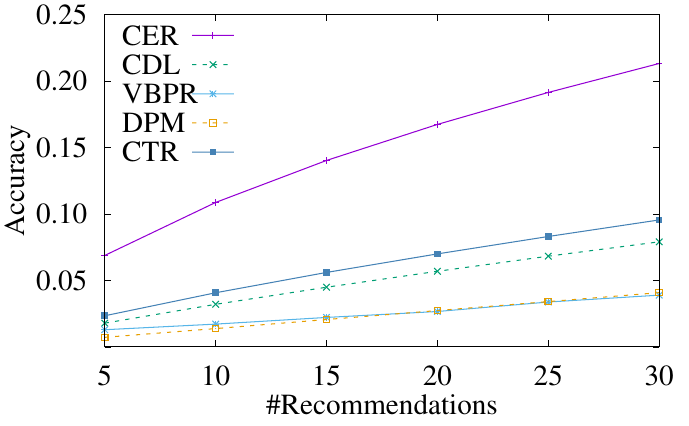}}
  \subfigure[CNNFV]{\label{fig:omcnnfvnf}\includegraphics[width=4cm]{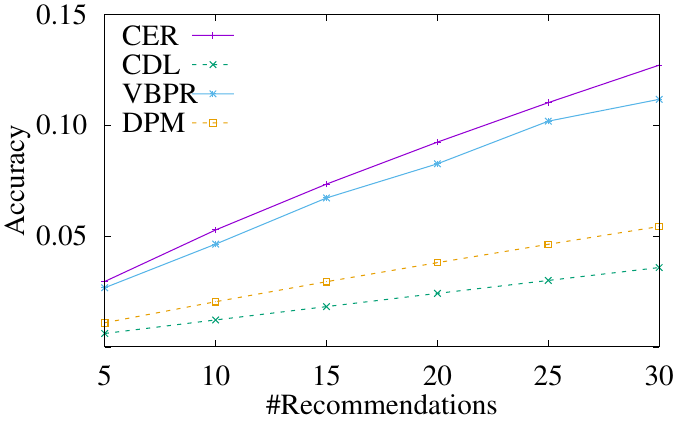}}
  \subfigure[WORD]{\label{fig:omwordnf}\includegraphics[width=4cm]{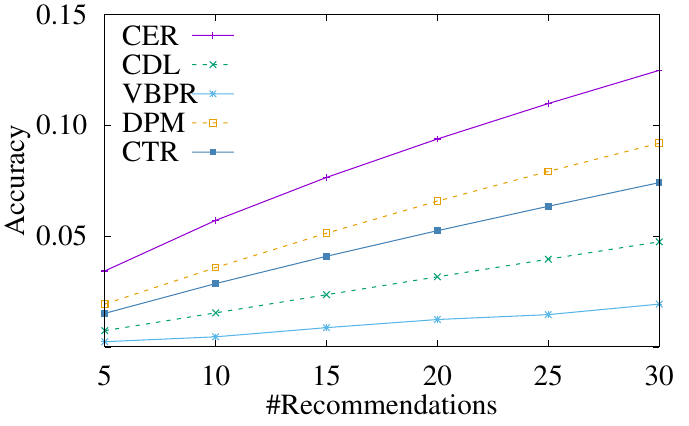}}
  \subfigure[CNNVLAD]{\label{fig:omcnnvladnf}\includegraphics[width=4cm]{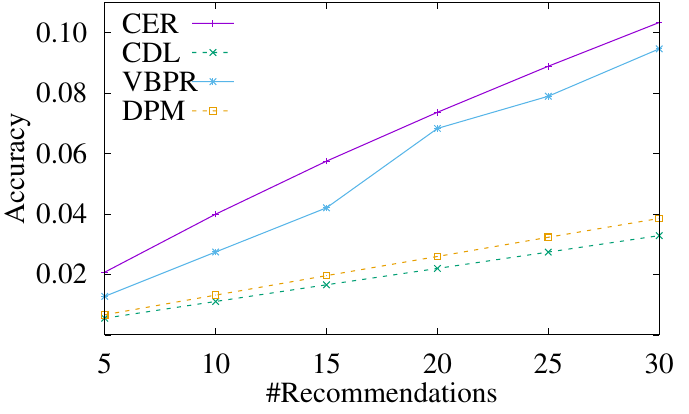}}
  \subfigure[IDT]{\label{fig:omidtnf}\includegraphics[width=4cm]{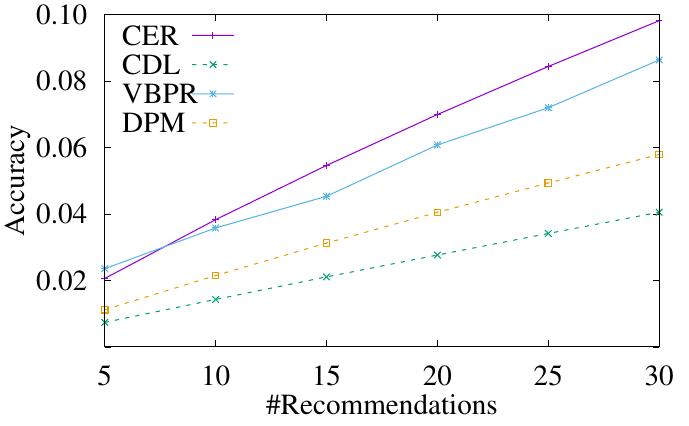}}
  \subfigure[MFCC]{\label{fig:ommfccnf}\includegraphics[width=4cm]{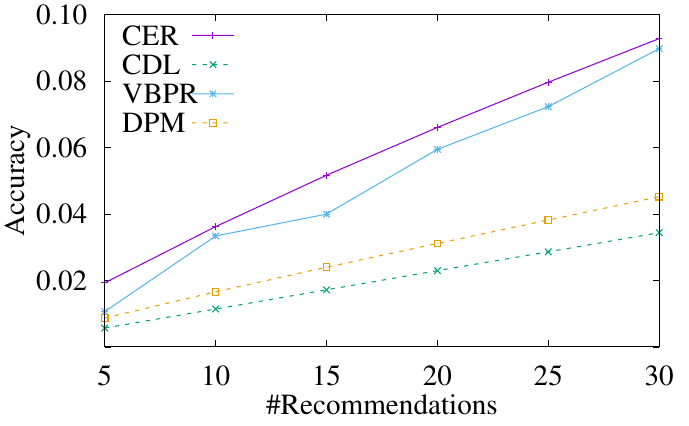}}
  \subfigure[MoSIFT]{\label{fig:ommosiftnf}\includegraphics[width=4cm]{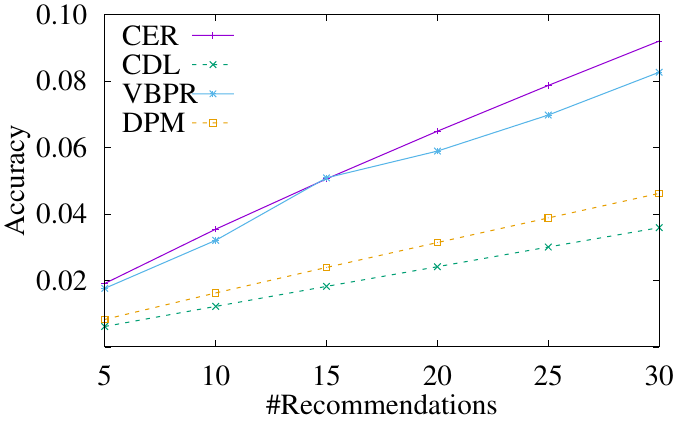}}
  \subfigure[OSIFT]{\label{fig:omosiftnf}\includegraphics[width=4cm]{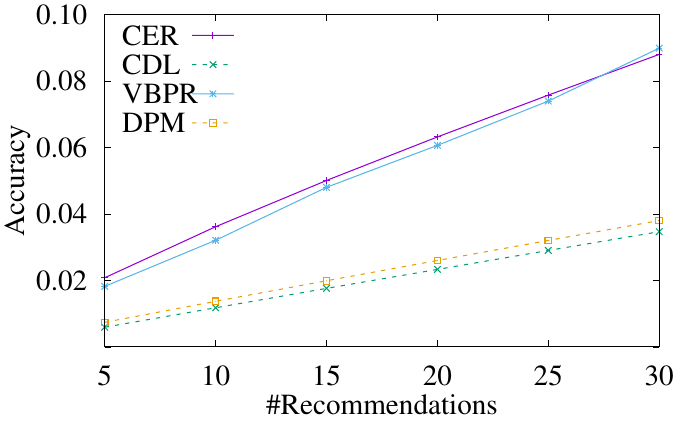}}
  \vspace{-10pt}
  \caption{Accuracy@k of different recommender methods and features in out-of-matrix setting on Netflix implicit dataset.}
  \label{fig:omaccnf}
  \vspace{-15pt}
\end{figure*}

In the out-of-matrix evaluation, we study two problems as follows:
\begin{itemize}
  \item How CER performs against other hybrid models with single content features in out-of-matrix scenario?
  \item Whether CER can generate accurate recommendations with arbitrary single content features in out-of-matrix scenario?
\end{itemize}
Specifically, we measure the out-of-matrix accuracy of CER , CDL, VBPR, DPM, and CTR on MovieLens (Fig. \ref{fig:omaccml}) and Netflix (Fig. \ref{fig:omaccnf}) with all the single content features. To compare these models clearly, each sub-figure in Fig. \ref{fig:omaccml} and \ref{fig:omaccnf} draws the accuracy of different models with one specific content feature, and the sub-figures from the same dataset are sorted by the Accuracy@$30$ of CER in a descending order.

From Fig. \ref{fig:omaccml} and \ref{fig:omaccnf}, we observe that CER consistently outperforms the other hybrid models on both datasets with any content feature, which validates that \textbf{CER is better than other hybrid models with any single content feature in out-of-matrix recommendation}. The comparison between CER and the other WMF-based models shows that linearly embedding the content features to the content latent vectors seems to be more general and effective than the existing content-side learning in existing hybrid models. We also observe that VBPR that also employs linear embedding achieves worst performances on Netflix with textual content features. The major difference between CER and VBPR is that CER is based on WMF while VBPR is based on BPR in the collaborative-side learning, we thus think the effectiveness of linear embedding holds only if it combines with WMF.

Another observation from Fig. \ref{fig:omaccml} and \ref{fig:omaccnf} is about the different impacts of single content features in CER's out-of-matrix recommendation. In details, the descending order of content features ranked by the Accuracy@$30$ on MovieLens dataset is $META>WORD>CNNFV>CNNVLAD>IDT>MFCC>OSIFT>MoSIFT$, while the order on Netflix dataset is $META>CNNFV>WORD>CNNVLAD>IDT>MFCC>MoSIFT>OSIFT$.
According to these comparison chains above, META vectors are the most effective content feature for video recommendation. This is because META vectors contain a lot of precise information that do not often appear in the video clips directly such as casts, producers and published years, which are more relevant with users' preferences on the videos. However, the META vectors are unavailable for most of the videos on a general video website like Youtube, therefore the accuracy of META vectors is only the ideal accuracy that CER could achieve. In practice, WORD vectors are more easily obtainable as they are made up of the video titles and descriptions. We observe that the recommendation accuracy with WORD vectors is a little higher than the accuracy with CNNFV vectors on MovieLens, yet a little lower on Netflix. Considering the video titles and descriptions we collected exactly reflect the video contents, the comparison indicates that, with the generality of CER, the non-textual content features could be the effective alternatives to textual features in common cases.
\begin{figure}[h!]
  \centering
  \subfigure[MovieLens]{\includegraphics[width=4cm]{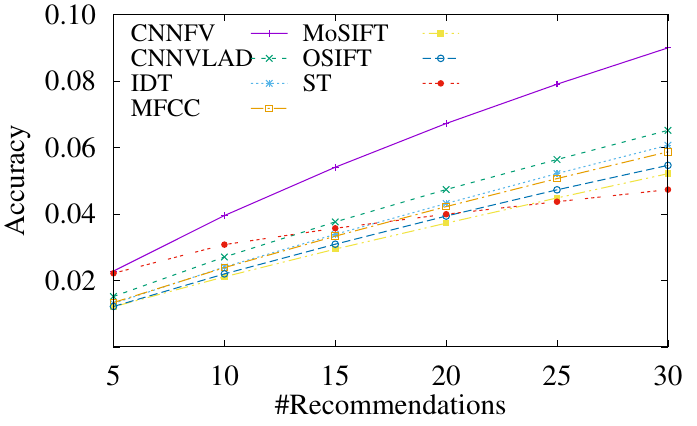}\label{fig:omstml}}
  \subfigure[Netflix]{\includegraphics[width=4cm]{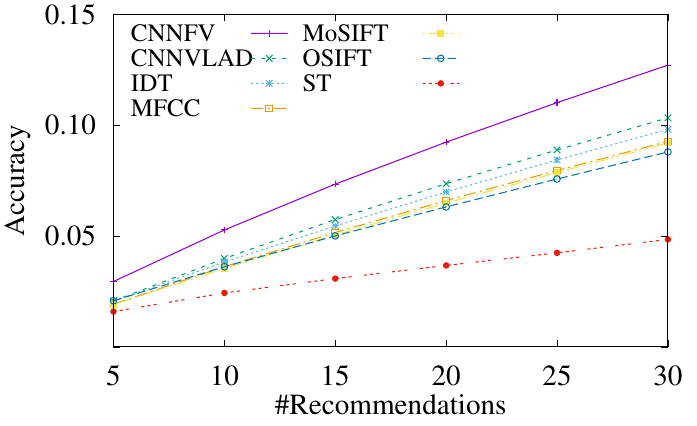}\label{fig:omstnf}}
  \vspace{-5pt}
  \caption{Out-of-matrix accuracy comparison in the sparse-text scenario}
  \label{fig:omst}
  \vspace{-10pt}
\end{figure}

To further examine the effectiveness of the non-textual part in our rich content features, we design the sparse-text WORD vectors (denoted as ST) that use only the video titles to perform the out-of-matrix recommendation again. The design bases on the fact that many user uploaded videos on Youtube with the titles only. We evaluate the accuracy of ST using CER and compare its accuracy with the non-textual content features in Fig. \ref{fig:omst}. The comparison shows that the CNNFV vectors significantly outperform the ST vectors on both datasets, while the other non-textual contents significantly outperform ST only on Netflix dataset. Putting the observations from Fig. \ref{fig:omst}, \ref{fig:omaccml}, and \ref{fig:omaccnf} together validates that, \textbf{CER is more general and effective than existing hybrid models in out-of-matrix scenario, although its accuracy varies significantly with different content features}.

\subsubsection{Efficiency of Model Training}
We also examine the time cost of training each recommender model. This experiment is conducted on two Intel E5-2690 v2 CPUs, one Nvidia Geforce 1070, and 256GB memory. All the WMF-based models are trained with CPU configuration, while all the BPR-based models are tested with GPU configuration (due to their slow training speeds in CPU configuration). We collect the time cost per iteration in the training stage of each model. Specifically, in WMF-based models, one iteration equals to update all the user latent vectors, all the video latent vectors, and the content side (if has) once. In the BPR-based models, one iteration needs to process all the training positives once. In addition, considering the textual content vectors are usually sparse and the non-textual content vectors are usually dense, we evaluate the time cost per iteration with both WORD vectors and CNNFV vectors.
\begin{table}[!hbt]
  \begin{center}
    \begin{tabular}{|>{\raggedright} p{30pt} |>{\raggedleft} p{40pt} |>{\raggedleft} p{30pt} |>{\raggedleft} p{40pt} |>{\raggedleft} p{30pt}|}
      \hline
      Content   & \multicolumn{2}{c|}{CNNFV}   & \multicolumn{2}{c|}{WORD} \tabularnewline \hline
      Dataset   & \centering MovieLens & \centering Netflix & \centering MovieLens & \centering Netflix \tabularnewline \hline
      CTR       & \multicolumn{2}{c|}{N/A}     & 15.46s  & 76.43s \tabularnewline \hline
      DPM       &   28.02s &   69.21s &   73.73s  & 94.32s \tabularnewline \hline
      CDL       &   45.37s &   82.43s &   175.82s & 141.10s \tabularnewline \hline
      VBPR      &   10.50s &   136.28s &   18.46s  & 204.82s  \tabularnewline \hline
      CER       &   11.00s &   67.10s &   14.62s  & 70.02s  \tabularnewline \hline \hline
      
      WMF       & MovieLens&   8.52s & Netflix  & 68.75s \tabularnewline \hline 
      BPR       & MovieLens&  8.10s & Netflix  & 116.08s \tabularnewline \hline
    \end{tabular}
  \end{center}
  \caption{The time cost per iteration in terms of seconds from each recommender model with two content features on two datasets.}
  \label{tbl:timecost}
  \vspace{-20pt}
\end{table}

Table~\ref{tbl:timecost} records the time cost of different models on both datasets. From Table \ref{tbl:timecost}, without the content-side learning, the BPR model costs less time per iteration than the WMF model on Movielens, while more time per iteration than WMF on Netflix. The reason behind this is that BPR's complexity is proportional to the number of ratings and WMF's complexity is proportional to the number of the users plus the number of the videos. When the training shifts from Movielens to Netflix, the number of training ratings increases faster than the total number of the users and the videos, which causes BPR to use more time on Netflix accordingly. Taking the content-side learning into consideration, CER achieves the lowest time cost per iteration in most cases. CER is more efficient than the other WMF-based models due to its lightweight content-side learning. In addition, CER is more efficient than VBPR in most cases although VBPR has GPU acceleration. This is because VBPR applies pair-wise learning that updates the video vectors double times compared to only once in the WMF-based models in each iteration. 

\begin{table*}[!t]
  \begin{center}
    \begin{tabular}{|>{\raggedright} p{50pt} |>{\centering} p{50pt} |>{\centering} p{50pt} |>{\centering} p{50pt} |>{\centering} p{50pt} |>{\centering} p{50pt} |>{\centering} p{50pt} |}
      \hline
      \multicolumn{7}{|c|}{Fusion on the rich non-textual content feature set} \tabularnewline\hline
      Method & Accuracy@5 & Accuracy@10 & Accuracy@15 & Accuracy@20 & Accuracy@25 & Accuracy@30 \tabularnewline \hline
      ACC & 0.022121 & 0.038951 & 0.053396 & 0.066657 & 0.078836 & 0.090326 \tabularnewline
      AVG & 0.021243 & 0.037600 & 0.051699 & 0.064683 & 0.076871 & 0.088040 \tabularnewline
      BPR & 0.021241 & 0.037602 & 0.051707 & 0.064690 & 0.076864 & 0.088044 \tabularnewline
      SVM & 0.021498 & 0.037748 & 0.051894 & 0.064918 & 0.077049 & 0.088096 \tabularnewline
      ECT & 0.021830 & 0.038222 & 0.051606 & 0.065720 & 0.077664 & 0.089460 \tabularnewline
      ESK & 0.021439 & 0.037826 & 0.051994 & 0.064998 & 0.077119 & 0.088350 \tabularnewline
      %0.024830,0.044222,0.060606,0.075720,0.089664,0.102460
      PRI (p=0.5) & \textbf{0.023090} & \textbf{0.040390} & \textbf{0.055746} & \textbf{0.069401} & \textbf{0.081788} & \textbf{0.093239} \tabularnewline \hline
      CNNFV & 0.022765 & 0.039240 & 0.053838 & 0.067042 & 0.078984 & 0.089915 \tabularnewline \hline \hline
      \multicolumn{7}{|c|}{Fusion on the rich content feature set} \tabularnewline\hline
      Method & Accuracy@5 & Accuracy@10 & Accuracy@15 & Accuracy@20 & Accuracy@25 & Accuracy@30 \tabularnewline \hline
      ACC & 0.064621 & 0.098163 & 0.124392 & 0.146985 & 0.166617 & 0.184319 \tabularnewline
      AVG & 0.063132 & 0.093990 & 0.118933 & 0.140554 & 0.159595 & 0.176996 \tabularnewline
      BPR & 0.061949 & 0.092322 & 0.116883 & 0.138128 & 0.156966 & 0.174059 \tabularnewline
      SVM & 0.067023 & 0.100991 & 0.127185 & 0.149112 & 0.168508 & 0.186059 \tabularnewline
      ECT & 0.041733 & 0.063908 & 0.081737 & 0.097422 & 0.111366 & 0.124557 \tabularnewline
      ESK & 0.068546 & 0.101244 & 0.127280 & 0.149542 & 0.169221 & 0.187168 \tabularnewline
      PRI (p=0.5) & \textbf{0.070307} & \textbf{0.105696} & \textbf{0.132879} & \textbf{0.156019} & \textbf{0.176204} & \textbf{0.194314} \tabularnewline \hline
      META & 0.065780 & 0.098771 & 0.124305 & 0.146079 & 0.165213 & 0.182318 \tabularnewline \hline
    \end{tabular}
  \end{center}
  \caption{Fusion accuracy of different methods on MovieLens implicit dataset with the rich content feature set and the rich non-textual content feature set}
  \label{tbl:fusionml}
  \vspace{-15pt}
\end{table*}

\begin{table*}[!t]
  \begin{center}
    \begin{tabular}{|>{\raggedright} p{50pt} |>{\centering} p{50pt} |>{\centering} p{50pt} |>{\centering} p{50pt} |>{\centering} p{50pt} |>{\centering} p{50pt} |>{\centering} p{50pt} |}
      \hline
      \multicolumn{7}{|c|}{Fusion on the rich non-textual content feature set} \tabularnewline\hline
      Method & Accuracy@5 & Accuracy@10 & Accuracy@15 & Accuracy@20 & Accuracy@25 & Accuracy@30 \tabularnewline \hline
      ACC & 0.029256 & 0.052103 & 0.072838 & 0.092046 & 0.110091 & 0.127141 \tabularnewline
      AVG & 0.028596 & 0.051106 & 0.071609 & 0.090616 & 0.108466 & 0.125376 \tabularnewline
      BPR & 0.028590 & 0.051106 & 0.071585 & 0.090590 & 0.108422 & 0.125329 \tabularnewline
      SVM & 0.028556 & 0.050811 & 0.070763 & 0.089278 & 0.106692 & 0.123241 \tabularnewline
      ECT & 0.029195 & 0.052478 & 0.073142 & 0.091845 & 0.100103 & 0.127278 \tabularnewline
      ESK & 0.028517 & 0.050983 & 0.071445 & 0.090408 & 0.108284 & 0.125155 \tabularnewline
      PRI (p=0.5) & \textbf{0.030394} & \textbf{0.054099} & \textbf{0.075226} & \textbf{0.094740} & \textbf{0.112876} & \textbf{0.130110} \tabularnewline \hline
      CNNFV & 0.029482 & 0.052753 & 0.073430 & 0.092353 & 0.110160 & 0.127079 \tabularnewline \hline \hline
      \multicolumn{7}{|c|}{Fusion on the rich content feature set} \tabularnewline\hline
      Method & Accuracy@5 & Accuracy@10 & Accuracy@15 & Accuracy@20 & Accuracy@25 & Accuracy@30 \tabularnewline \hline
      ACC & 0.072106 & 0.114397 & 0.148046 & 0.176939 & 0.202550 & 0.225676 \tabularnewline
      AVG & 0.065187 & 0.104734 & 0.136801 & 0.164413 & 0.189063 & 0.211580 \tabularnewline
      BPR & 0.063911 & 0.102926 & 0.134620 & 0.162095 & 0.186601 & 0.208990 \tabularnewline
      SVM & 0.052362 & 0.084992 & 0.112074 & 0.135954 & 0.157501 & 0.177489 \tabularnewline
      ECT & 0.039491 & 0.065230 & 0.087030 & 0.106488 & 0.124300 & 0.140920 \tabularnewline
      ESK & 0.066386 & 0.106135 & 0.138135 & 0.165710 & 0.190251 & 0.212697 \tabularnewline
      PRI (p=0.5) & \textbf{0.073158} & \textbf{0.115894} & \textbf{0.149731} & \textbf{0.178567} & \textbf{0.204068} & \textbf{0.227094} \tabularnewline \hline
      META & 0.068739 & 0.108569 & 0.140139 & 0.167226 & 0.191242 & 0.213063 \tabularnewline \hline
    \end{tabular}
  \end{center}
  \caption{Fusion accuracy of different methods on Netflix implicit dataset with the rich content feature set and the rich non-textual content feature set}
  \label{tbl:fusionnf}
  \vspace{-30pt}
\end{table*}

\subsection{Evaluation on Multiple Feature Fusion}
In this experiment, we explore whether the proposed priority-based late fusion method (PRI) can make the recommendation more accurate with multiple content features. Our fusion evaluation is based on CER, because it is the most accurate method in our single content evaluation. We prepare the rich content features and the rich non-textual content features for CER, and apply different fusion methods with CER's model or outputs. The accuracy values are measured with out-of-matrix scenario where multiple feature fusion should be effective according to the existing works \cite{srivastava2012learning} \cite{aly2013axes}. Table \ref{tbl:fusionml} and \ref{tbl:fusionnf} record the accuracy results of different fusion methods on MovieLens and Netflix respectively. In these tables, we use the accuracy results from the best single content feature in the corresponding rich content features as the baseline to help judge whether a fusion method improves the accuracy actually. In addition, we highlight the results of the proposed late fusion method (PRI) with the bold font.

From Table \ref{tbl:fusionml} and \ref{tbl:fusionnf}, we firstly check the out-of-matrix accuracy results of the early fusion methods. We observe that ECT fusion's accuracy significantly drops when it is applied on the rich content features. This observation shows that concatenating the heterogeneous content vectors together to learn the shared space harms the recommendation accuracy in the video recommendation scenario. Compared to ECT, ESK that adds the content latent vectors in the shared space achieves consistent recommendation accuracy in out-of-matrix scenario, and beat ACG, BPR and SVM most of the time. However, both of the early fusion methods fail to beat the baseline all the time in our evaluation. These comparisons validate that the heterogeneous content vectors are hard to form more meaningful shared space in the video recommendation scenario.

\begin{figure}[h!]
	\centering
	\subfigure[MovieLens]{\includegraphics[width=4cm]{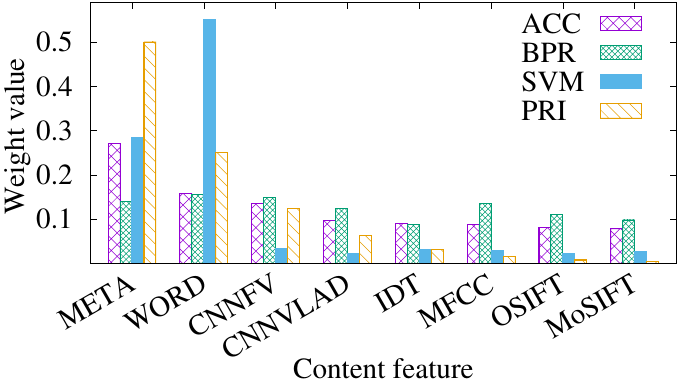}\label{fig:wbml}}
	\subfigure[Netflix]{\includegraphics[width=4cm]{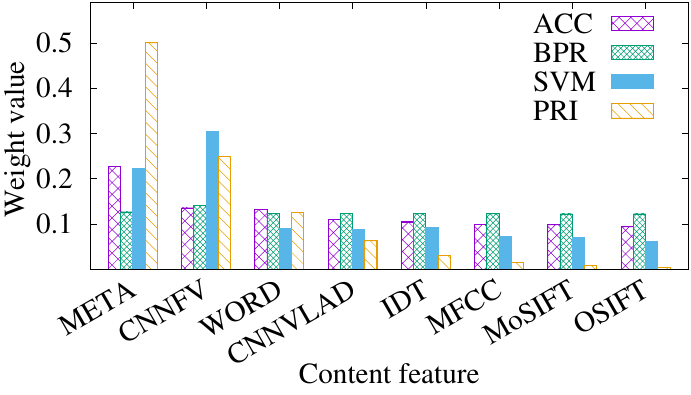}\label{fig:wbnf}}
  \vspace{-10pt}
	\caption{Weight distributions from different late fusion methods}
	\label{fig:wb}
	\vspace{-10pt}
\end{figure}

The early fusion methods' robustness and effectiveness problems also exist in the late fusion methods. In details, AVG, BPR and SVM only beat the baseline once when they are applied with rich content features on MovieLens. The possible failure of AVG is due to the huge performance gaps among different content features. In Fig. \ref{fig:omaccml} and \ref{fig:omaccnf}, the highest out-of-matrix accuracy of CER is achieved with META vectors, while the lowest accuracy of CER is achieved with MoSIFT vectors. The highest is three times to the lowest. As a result, average fusion (AVG) weakens the predictability of more powerful content vectors. Both BPR and SVM are learning-to-rank methods. They learn content weights by regarding the predicted ratings from contents as the feature vectors. Nevertheless, these existing learning-to-rank methods cannot learn the content priorities properly in our evaluation. The evidences are shown in Fig. \ref{fig:wb} where we draw weight distributions over all the content features of each late fusion method after normalising the weight sum to $1$. In Fig. \ref{fig:wb}, if we regard the weights from ACC as the correct ones, those from BPR and SVM are wrong: BPR treats different contents almost equally, while SVM cannot correctly lift the weight of META over that of WORD according to their significant highest out-of-matrix accuracy differences. Eventually, they fail to generate more accurate recommendations. Despite PRI, our heuristic late fusion method -- ACC achieves the largest improvement in Table \ref{tbl:fusionml} and \ref{tbl:fusionnf}. It always beats the baseline, which initially shows the content priorities in late fusion is necessary.

The proposed PRI achieves the highest accuracy in our evaluation, and it beats the baseline on both datasets all the time. Compared to all the other fusion methods, we think the success of PRI has two reasons. First, PRI prioritizes content features as ACC does. This makes PRI beats the other comparison methods as ACC does. Second, PRI does not simply trusts the content features by their accuracy values as ACC does. Accordingly, as Fig. \ref{fig:wb} shows, PRI assigns larger weights to the more powerful content features to protect their priorities in the fusion. This makes PRI further beats ACC. It is worth noting that PRI's improvement only happens on CER. For other models, because the absolute performance divergences are small, PRI could not improve the accuracy further even though rich content set presents (due to page limitation, we do not put PRI's results with other models in this paper). The possible reason is that CER's large absolute performance divergences between different features indicate correct feature priorities in out-matrix scenario. In summary, our evaluation validates that \textbf{PRI can improve the video recommendation accuracy further when large absolute performance divergences between individual features exist}.

\section{Conclusion}

In this paper, we explored how to exploit rich content features from videos to improve personalized recommendation accuracy in both in-matrix and out-of-matrix scenarios. We assumed that the rich content features can improve the hybrid recommender methods in two situations: (1) when single specific content features are unavailable; (2) when multiple content features are available. To validate our assumptions, we firstly extracted multiple textual and non-textual content features from videos. Our initial evaluation showed that existing hybrid models were only effective in either in-matrix or out-of-matrix scenario with rich content features. To overcome the limitation, we proposed a general and effective hybird model, namely collaborative embedding regression (CER), to marginalize the performance drop caused by the unavailability of one specific content. In addition, we studied how to fuse multiple heterogeneous content features to improve the recommendation accuracy further. We proposed a priority-based late fusion method (PRI) to effectively fuse both non-textual and textual content features. We conducted extensive evaluations on single content features to validate the effectiveness of CER, and evaluations on multiple content features to validate the effectiveness of PRI. The experimental results on MovieLens and Netflix datasets showed that the proposed CER and PRI are superior to existing hybrid recommender models and multiple feature fusion methods respectively. In particular, CER has more significant impact on the recommendation accuracy and can work well with existing late fusion methods. PRI must work with the models which has quite diverse performance on different content features.  In summary, all the experiments validated our initial assumptions and revealed the benefits of the rich content features for personalized video recommendation.

\bibliographystyle{IEEEtran}
\bibliography{VCRS}

\end{document}